\documentclass[a4paper,fleqn]{cas-sc}

\usepackage[authoryear]{natbib}
\usepackage{subfigure}
\usepackage{amsmath}
\usepackage{bm}
\usepackage[T1]{fontenc}
\usepackage{graphicx}
\usepackage{color}
\usepackage{hyperref}

\def\tsc#1{\csdef{#1}{\textsc{\lowercase{#1}}\xspace}}
\tsc{WGM}
\tsc{QE}
\tsc{EP}
\tsc{PMS}
\tsc{BEC}
\tsc{DE}

\begin{document}
\let\WriteBookmarks\relax
\def\floatpagepagefraction{1}
\def\textpagefraction{.001}
\shorttitle{Microstructural origins of crushing strength}
\shortauthors{D Cantor et~al.}

\title [mode = title]{Microstructural origins of crushing strength for inherently anisotropic brittle materials}

\author[1,2]{David Cantor}[orcid=0000-0003-3354-1818]
\cormark[1]
\ead{david.cantor@tpolymtl.ca}

\author[1,2]{Carlos Ovalle}[orcid=0000-0002-9648-5262]
\ead{carlos.ovalle@polymtl.ca}

\author[3,4]{Emilien Az\'ema}[orcid=0000-0001-8831-3842]
\ead{emilien.azema@umontpellier.fr}

\address[1]{Department of Civil, Geological and Mining Engineering, Polytechnique Montreal, Montreal, QC, Canada H3T1J4}
\address[2]{Research Institute of Mining and Environment, RIME UQAT-Polytechnique, Montreal, QC, Canada H3T1J4}
\address[3]{LMGC, Universit\'e de Montpellier, CNRS, Montpellier, France 34090}
\address[4]{Institut Universitaire de France (IUF), Paris, France}

\cortext[cor1]{Corresponding author}

\begin{abstract}
We study the crushing strength of brittle materials whose internal structure (e.g., mineral particles or graining) presents a layered arrangement reminiscent of sedimentary and metamorphic rocks. 
Taking a discrete-element approach, we probe the failure strength of circular-shaped samples intended to reproduce specific mineral configurations. 
To do so, assemblies of cells, products of a modified Voronoi tessellation, are joined in mechanically-stable layerings using a bonding law. 
The cells' shape distribution allows us to set a level of inherent anisotropy to the material. 
Using a diametral point loading, and systematically changing the loading orientation with respect to the cells' configuration, we characterize the failure strength of increasingly anisotropic structures. 
This approach ends up reproducing experimental observations and lets us quantify the statistical variability of strength, the consumption of the fragmentation energy, and the induced anisotropies linked to the cell's geometry and force transmission in the samples. 
Based on a fine description of geometrical and mechanical properties at the onset of failure, we develop a micromechanical breakdown of the crushing strength variability using an analytical decomposition of the stress tensor and the geometrical and force anisotropies. 
We can conclude that the origins of failure strength in anisotropic layered media rely on compensations of geometrical and mechanical anisotropies, as well as an increasing average radial force between minerals indistinctive of tensile or compressive components. 
\end{abstract}




\maketitle

\section{Introduction}
The mechanical behavior of many solids can be tracked down to the level of molecules, defects, and dislocations that may cause stress concentrations and yielding of the material. 
However, the microstructural level can be equally or more important for characterizing their failure strength. 
Imagine, for instance, arrangements produced by the genesis, layering, and stratification of a solid such as sedimentation, rock metamorphism, or even crystal growth. 
In these cases, the mechanical properties are not determined at the molecular level but instead at the microstructural level \citep{Griffith1921}. 

We focus on the failure strength of brittle materials, which may include rocks, soil grains, ceramics, and even ice. 
In the case of rock masses and rock aggregates, for example, it is well known that the mechanical behavior depends on mineralogy and graining characteristics (e.g., size and shape distribution of minerals), matrix level of cementation, diagenesis, joint characteristics, and fissuring \citep{Jaeger2007}.
Many of these geometrical attributes are also called \emph{fabric} or \emph{microstructure} for geological materials. 
If any of those fabric properties present a preferred orientation or organization in space, then the material can be considered inherently anisotropic. 
Many studies have focused on the quantification of the level of inherent fabric anisotropy in rocks or on the impact of the loading orientation on the ultimate strength of such layered materials \citep{Hoek1964,Oda1982,Amadei1996,Chen1998,Zhang2000,Karakul2010, Khanlari2015, GuhaRoy2016,Pouragha2020}. 

\begin{figure}
    \centering
    \includegraphics[height=0.25\linewidth]{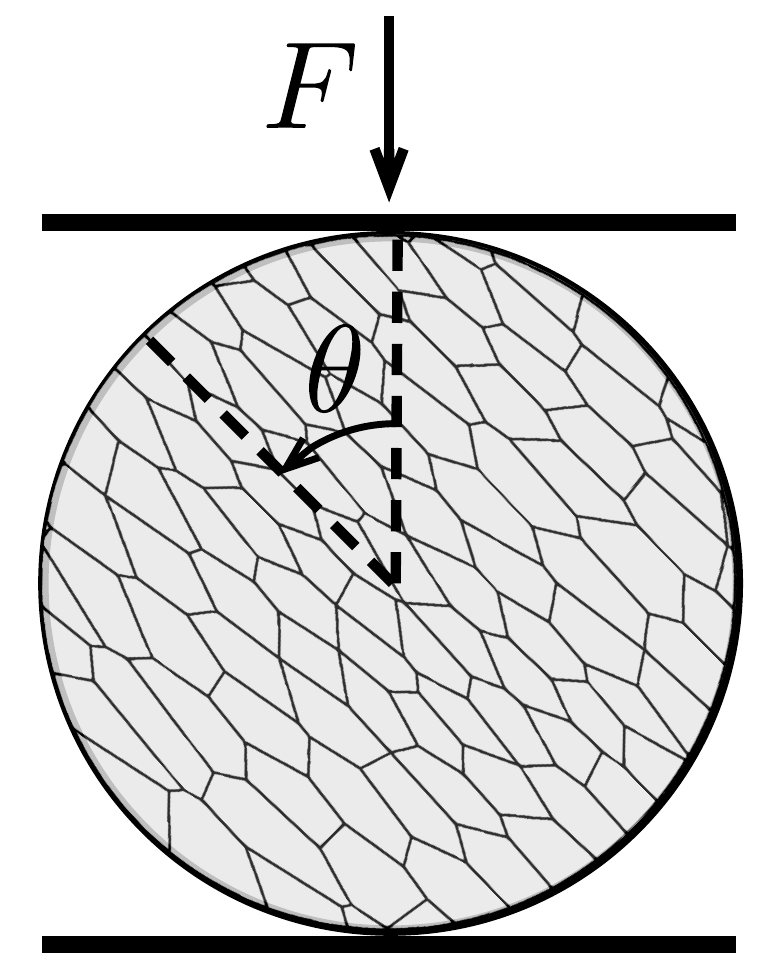}
    \caption{Scheme of a diametrical loading in which the loading orientation varies relative to the orientation of the internal structure.}
    \label{fig:load_scheme}
\end{figure}

In experiments, cylindrical cores are often used to characterize the failure strength of inherently anisotropic rocks under diametrical point loading (commonly called \emph{Brazilian test}). 
In those tests, the orientation $\theta$ of the applied force is gradually varied with respect to the orientation of the internal layering (see Fig. \ref{fig:load_scheme}). 
For such a circular geometry, axial symmetry is found along the layering orientation; then, the failure strength can be fully characterized by varying $\theta$ in the range $[0^{\circ}, 90^{\circ}]$. 

For rocks not presenting an inherent anisotropy, the failure strength is independent of the loading orientation $\theta$. 
However, for highly layered rocks such as slate, schist or shale, the failure strength largely varies with $\theta$ in a `\emph{U}' shape with minimal strength for an orientation around $\theta \simeq 25^{\circ}$, and increasing strength as $\theta \rightarrow 0^{\circ}$ or $90^{\circ}$ \citep{Amadei1996,Saroglou2008,Karakul2010,Saeidi2014,Garagon2010,Pouragha2020,Xu2020}.
As shown early by \citeauthor{Hoek1964} in 1964, the `\emph{U}' shape is consistent with \citeauthor{Griffith1921}'s theory of brittle fracture of materials, in which the maximum stress at the tip of the crack triggers the propagation of a fissure once a critical amount of energy is added to the system.
The rock microstructure is then capable of signing the failure modes and patterns.
This in turn significantly affects the stress-strain relations at a macroscopic level (i.e., stiffness, hardening/softening, and strength) \citep{Hurley2018,Marinelli2019}. 

It is important to clearly distinguish between inherent and induced anisotropies. 
While the first is defined here as a property of the fabric, the latter refers to anisotropies arising from that primary structure, such as the joint distribution in space. 
Although the anisotropies mentioned up this point are only related to geometrical properties, they can also arise from the loading configuration (i.e., stress-induced anisotropies). 
Indeed, a detailed description of both inherent and induced anisotropies are key elements for understanding the behavior of brittle materials, as we will show in this paper. 

Improving our understanding of anisotropic geological materials will help us better address problems involving rock and grain fragmentation, such as railway ballast design \citep{Indraratna2011,Lim2004}, rockfill dam design \citep{Marsal1973,Ovalle2014,Ovalle2020}, rock tunneling processes \citep{Favier2006,Pindra2010}, mining waste dumps construction \citep{Bard2013}, surface subsidence \citep{Brzesowsky2014}, slip stability analysis of fault gouges \citep{Main1991}, filling rock mass discontinuities \citep{Sammis1987}, the geological formation of glacial till \citep{Turcotte1986}, confined comminution \citep{Einav2007,Ovalle2016}, weathering and environmental degradation effects \citep{Zhang2018}, etc. 

In this paper, we use bi-dimensional discrete-element modeling to study the failure strength of circular samples that have an inherent anisotropic configuration under varied loading orientation. 
In Sec. \ref{Sec:NumModel}, we introduce our numerical strategy based on the contact dynamics and the bonded-cell methods, and the sample construction and testing procedures. 
In Sec. \ref{Sec:Macro}, we characterize the failure strength of inherent anisotropic structures showing a good agreement with experimental observations. 
We then analyze the failure strength variability in terms of Weibull's statistics and failure mode evolution. 
Section \ref{Sec:Micro} focuses on a fine description of the microstructure in terms of fabric connectivity, force transmission, and inherent and induced anisotropies. 
In Sec. \ref{Sec:Model}, we develop a theoretical analysis that allows us to discover the microstructural origins of the crushing strength in terms of the level of inherent anisotropy and loading orientation. 
This analytical approach based on the granular stress tensor and its harmonic decomposition linking microstructure and the macromechanical response. 
Finally, we conclude with a summary and perspectives. 

\section{Numerical modeling}\label{Sec:NumModel}
Inherently anisotropic materials are challenging to characterize given the complex and multiscale properties that minerals, graining, bonds, and fissures can present in space. 
Numerical approaches have proven successful at analyzing these materials because they are able to reproduce complex failure mechanisms under controlled geometries. 
Some of these approaches use, for example, finite-elements \citep{Sulem2002,Beyabanaki2009}, discrete-element methods with bonded bodies \citep{Potyondy2004,Cho2007,Lan2010,Scholtes2013,Kazerani2013,Gao2014}, splitting or replacing mechanisms \citep{Cantor2015,Ciantia2015,Gladkyy2017,Iliev2019}, or coupled discrete-finite element strategies \citep{Mahabadi2010,BagherzadehKh2011,Guo2014,Ma2014}.

Among these approaches, the discrete-element method (DEM) has become increasingly popular for dealing with fragmentation due to its versatility in reproducing grain fissuring, crushing, and many experimental observations. 
However, some modeling strategies employ circular particles to represent grains and blocks, which does not capture the complex variability of fragments' shapes and sizes. 
Other studies use a `replacement' method in which bigger grains are substituted by a set of smaller bodies once a criterion is reached, at the expense of missing mass conservation or creating local over-stresses at the replacement instant. 
Finally, energy consumption is not traceable when using circular bodies, or \emph{ad-hoc} parameters are necessary to estimate the fragmentation energy. 
While these approaches have enabled the exploration of certain mechanisms of rock and grain failure, no clear mapping between the variability of strength, failure modes, and the microstructure has been found for inherently anisotropic materials. 
A correct simulation of these materials requires a model in which bodies can break into irregular and size disperse fragments while simultaneously controlling the inherent anisotropy level. 
As we show in the next section, these conditions can be met in 2D simulations using irregular convex polygons. 

\subsection{Construction of inherently anisotropic samples}
We build circular samples composed of smaller bodies called \emph{cells} using a Voronoi decomposition of a unitary circle. 
This procedure generates an assembly of $N_{cl}$ adjacent cells that we `glue' using a cohesive bonding law. 
This approach, known as the bonded-cell method (BCM), has been used in numerous studies of the mechanical behavior of crushable granular materials, both in 2D \citep{Nguyen2015} and 3D \citep{Cantor2016,Orozco2019,Huillca2020}. 

\begin{figure}
    \centering
    \includegraphics[width=0.72\linewidth]{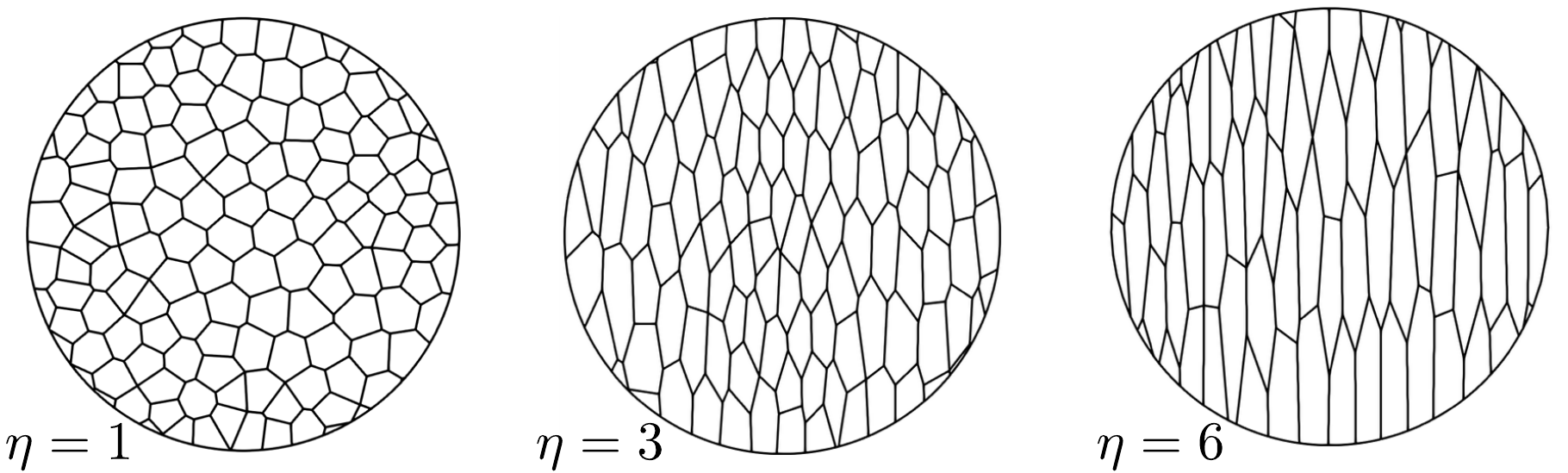}
    \caption{Samples presenting increasing average cell aspect ratio $\eta$. We varied $\eta$ from $\mathrm{(a)} \ 1$ up to $\mathrm{(f)} \ 6$ in steps of 1.} 
    \label{fig:example_part}
\end{figure}

A random Voronoi tessellation normally creates a disordered distribution of cell shapes and sizes. 
In order to control the cell's geometry (and, in effect, the inherent anisotropy), we alter the initial tessellation in two steps. 
First, we iteratively rebuild the Voronoi tessellation using the centroids of previous tessellation seedings to produce similar cells.
This approach is also called centroid tessellation \citep{Du2006}. 
Then, the cells are elongated along a given direction and an anisometry level is estimated using the average aspect ratio of the cells $\eta = h/L$, with $h$ and $L$ being the average short and long dimensions, respectively. 
This anisometry represents the inherent anisotropy configuration of the minerals in our model. 
We produced a set of samples with $\eta = [1,6]$ in steps of 1 (see Fig. \ref{fig:example_part}). 
Additionally, perturbations to the initial setting of the tessellation enabled us to have slightly different cell arrangements. 
For statistical representativeness, we built five different configurations for each value of $\eta$. 

In order to give mechanical strength to the assembly of cells, we define a normal and tangential cohesion at the bonds (i.e., cell-cell interactions), $C_n$ and $C_t$, respectively. 
$C_n$ prevents the interactions between cells from separating due to tensile stresses, while $C_t$ provides resistance against sliding. 
We also preset a debonding distance $\delta_c$ needed to effectively break a cohesive bond. 
By choosing a typical value of surface energy density for silicate minerals $\gamma = 50 \ \mathrm{J/m^2}$ \citep{Jones2019}, we can then determine the separation threshold as $\delta_c = 2\gamma/C_n$, following fracture mechanics theory. 
Note that our model allows us to independently define the tensile and shear bonding strength, but for simplicity we set $C_n = C_t$. 
A detailed analysis of the combined effect of varying $C_n$ and $C_t$ can be found in Ref. \cite{Cantor2016}. 

\begin{figure}
    \centering
    \includegraphics[height=0.25\linewidth]{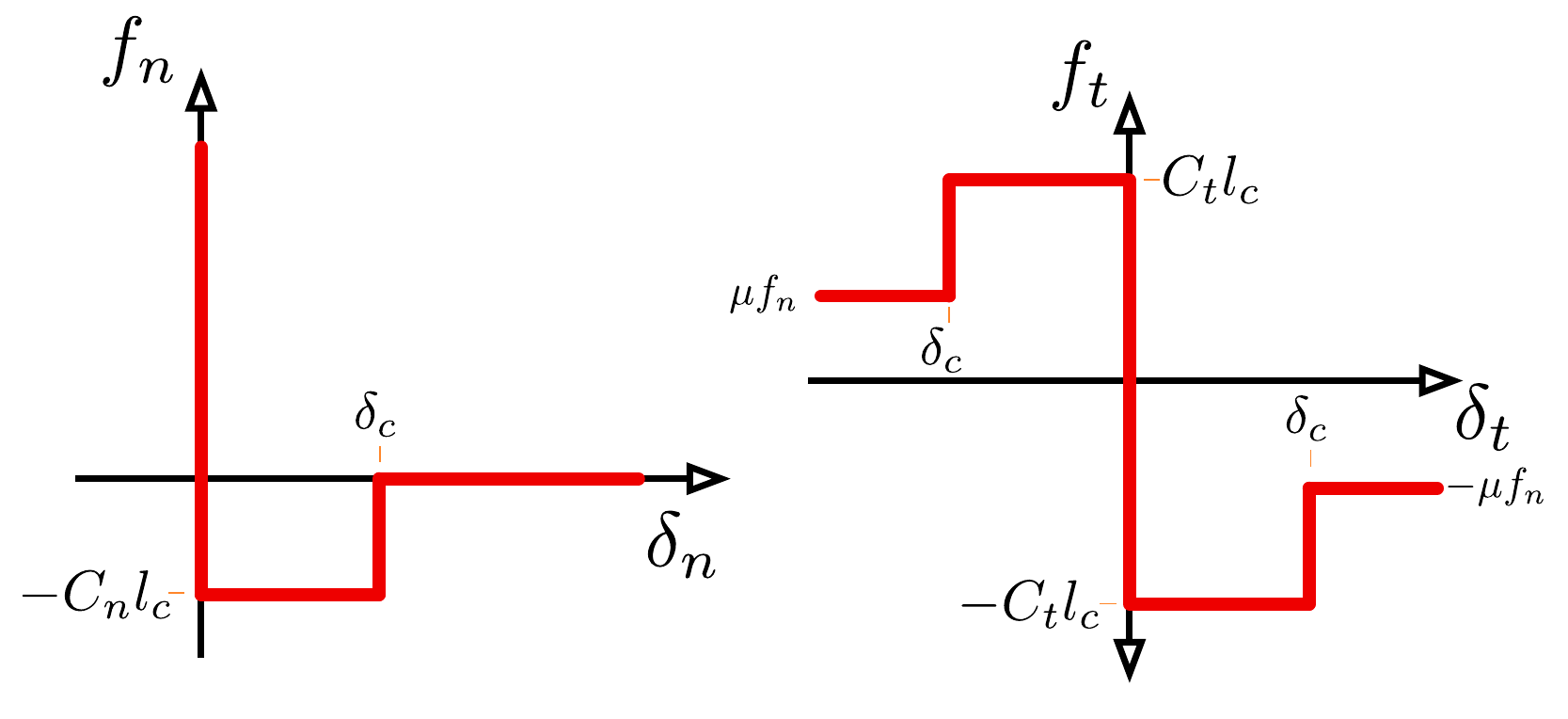}
    \caption{Interaction law for cohesive bonds between cells for the normal (left) and the tangential components (right) relative to the local framework coordinates.}
    \label{fig:ctc_law}
\end{figure}

The critical rupture energy that a bond needs to break is thus $E_c = 2 \gamma l_c$, with $l_c$ the length of the interaction.
Once $E_c$ is reached, the cohesive bond is removed, simulating a fissuring event. 
These fissures are considered dry frictional surfaces, with $\mu$ being the coefficient of friction that we set to $0.4$ (see Fig. \ref{fig:ctc_law} for a schematic representation of the bonding law).

In addition, numerical studies have explored the effect of the number of cells on the failure strength of brittle materials, showing that an increased number of cells lowers the failure strength \citep{Nguyen2015,Huillca2020}. 
However, it was recently shown that the scalability of failure strength is not simply linked to the number of cells, but more importantly to the length of bonding interactions \citep{Orozco2019,Cantor2021}. 
Thus, to make the tests comparable, samples must present the same potential surface energy among the different values of $\eta$ despite presenting a different number of cells. 
So, in our tests, the samples have the same total length of bonds. 


\subsection{Contact dynamics}
The contact dynamics (CD) method is a discrete-element approach in which rigid bodies interact via \emph{non-smooth} laws \citep{Jean1999,Dubois2018}, i.e., impacts are transmitted on an implicit time-stepping scheme. 
At the end of each time-step, particles' velocities and contact forces can therefore be simultaneously computed without requiring force-overlapping laws (i.e., no regularization of the contact law is needed). 
This allows the CD method to be unconditionally stable and capable of employing larger time-steps than in alternative \emph{smooth} approaches. 
For details on implementation of the contact dynamics method, see Refs. \citep{Renouf2004,Radjai2009}.

\begin{figure}
    \centering
    \includegraphics[height=0.2\linewidth]{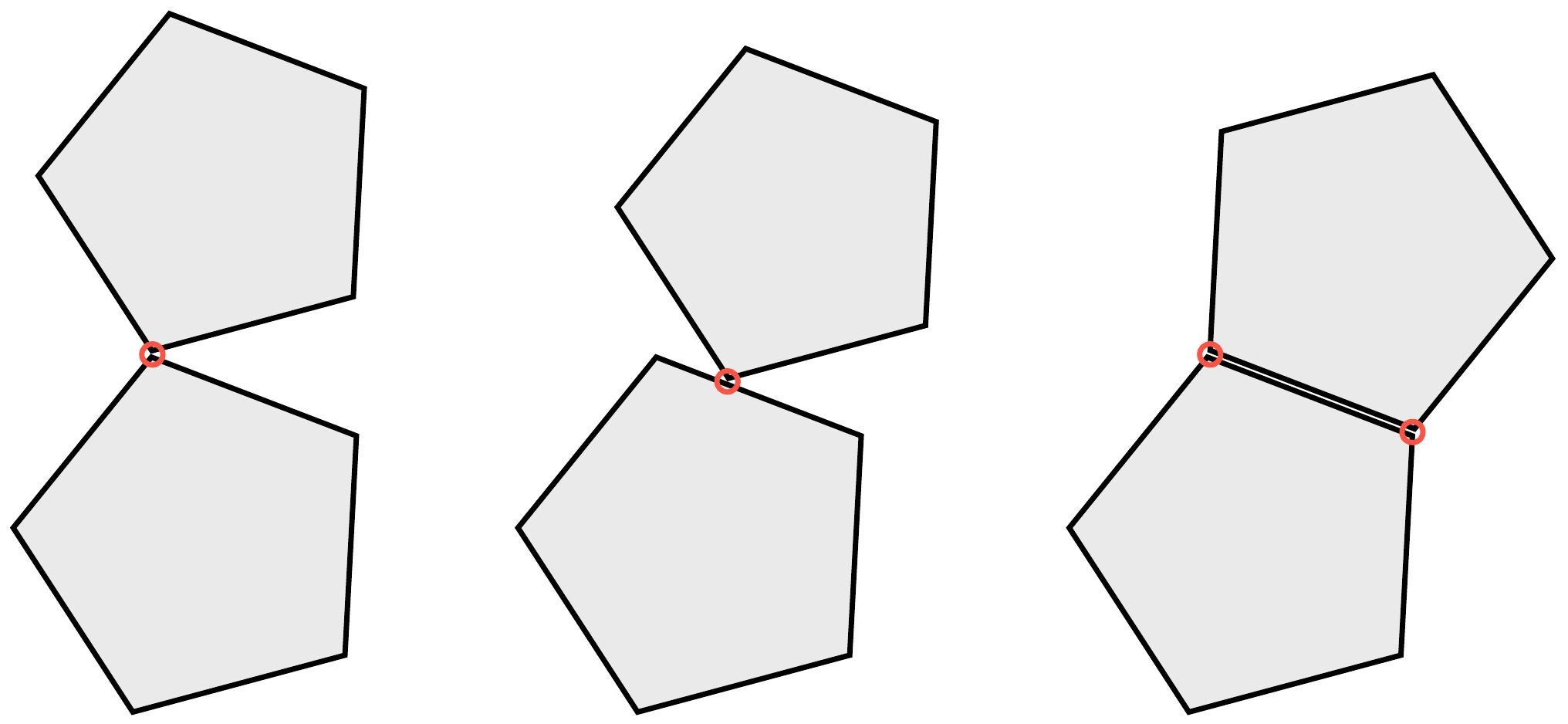}
    \caption{Different interaction types between convex polygons: (left) vertex-vertex, (center) vertex-edge, and (right) edge-edge. }
    \label{fig:ctc_type}
\end{figure}

In two-dimensional simulations, three main interactions can occur between convex bodies: vertex-vertex, vertex-edge, edge-edge (see Fig. \ref{fig:ctc_type}). 
Vertex-vertex interactions are rare and unstable, so they are discarded from the computation and analysis. 
For edge-edge interactions, it is necessary to consider two contact points to correctly resolve the contact mechanics; however, only the resultant force is important, rendering the loci of the two contact points irrelevant. 
For the interaction detection and classification, we use the \emph{shadow-overlap method} \citep{Saussine2006}, which creates a separating plane between two touching bodies via an iterative procedure. 
Updated body positions, velocities, and interaction forces are governed by the equations of motion and the cohesive bonding law we previously defined. 

Finally, it is worth mentioning that our simulations were performed using the CD method on the free and open-source platform LMGC90 \citep{Dubois2011,LMGC90Web}. 
\subsection{Test procedures}
Samples are initially set matching the loading orientation with the cells' preferred direction, so $\theta = 0^{\circ}$.
Then, we apply a gradually increasing vertical force $F$ using rigid platens up to the failure. 
To avoid dynamic perturbations during loading, we make sure that, over a time-step, a load increment is very small before $C_n d$, with $d$ being the diameter of the samples. 
We systematically vary $\theta$ in the range $[0^{\circ},90^{\circ}]$ in steps of $5^{\circ}$, i.e., 18 different orientations (see Fig. \ref{fig:load_scheme}). 
Finally, we test five different configurations of $\eta$ for each one of the angles for statistical representativeness. 
Videos of the tests can be found at the following link \href{https://youtu.be/-houBSR4b2A}{https://youtu.be/-houBSR4b2A}.
We present the averaged results of the total 540 simulations we performed. 

\section{Macroscopic observations}\label{Sec:Macro}
\subsection{Failure strength}
\begin{figure}
    \centering
    \subfigure [ ]{\label{fig:brittle}
    \includegraphics[width=0.3\linewidth]{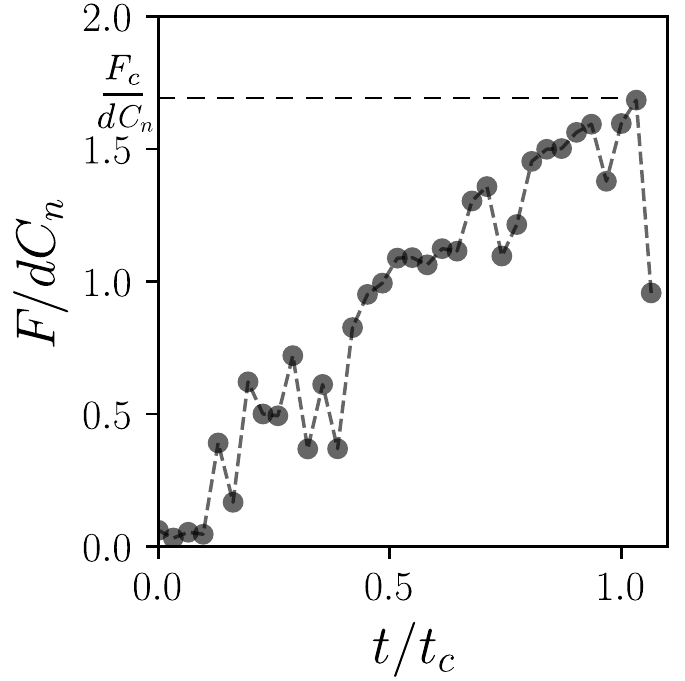}}
    \subfigure [ ]{\label{fig:crushing_strength}
    \includegraphics[width=0.5\linewidth]{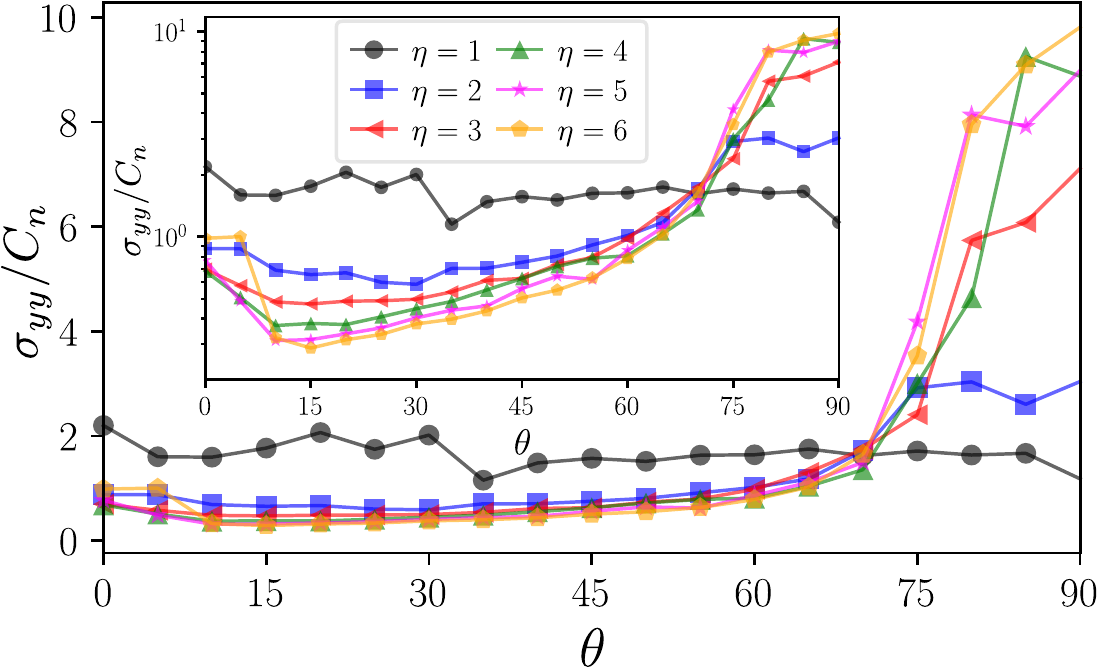}}
    \caption{(a) Typical evolution of the reaction force $F$ on the loading plate as a function the simulation time normalized by the time at failure $t_c$. (b) Strength $\sigma_{yy}$ normalized by the normal bonding cohesion $C_n$ as a function of the relative loading orientation $\theta$ and different levels of inherent anisotropy $\eta$. 
    In the inset, we present the same data in lin-log scale.}
    \label{fig:crushing_strength_all}
\end{figure}

Our samples are able to reproduce a brittle material behavior by supporting a load that gradually increases up to a critical value that triggers the collapse of the assembly (see Fig. \ref{fig:brittle}).
The critical force at failure $F_c$ allows us to characterize strength using the vertical stress at failure, defined as 
\begin{equation}\label{eq:sigma_yy}
\sigma_{yy} = \frac{F_c}{d}. 
\end{equation}

Figure \ref{fig:crushing_strength} summarizes the average values for $\sigma_{yy}$ found in our tests as a function of $\theta$ and the inherent anisotropy $\eta$. 
Under the same loading rates, the internal cohesion is a natural scaling parameter for these systems. 
So, we plot $\sigma_{yy}$ normalized by $C_n$.

Note that for the case $\eta=1$, in which the cells do not present any characteristic orientation, the strength is independent of the angle $\theta$. 
For the anisotropic configurations where $\eta>1$, the strength remains relatively similar for loading orientations below $\theta \simeq 70^{\circ}$, but always underneath the values found for the case $\eta=1$.
Beyond $\theta \simeq 75^{\circ}$, there is an important gain in strength, which seems accentuated as $\eta$ increases. 
Finally, a maximum failure strength is found for loading orientation perpendicular to the layering of the cells (i.e., $\theta = 90^{\circ}$). 

As previously mentioned, experimental observations have systematically highlighted the `\emph{U}' shape displayed in Fig. \ref{fig:crushing_strength}, with a critical loading orientation $\theta_c$ exhibiting the minimal strength. 
Simple stress considerations can predict that critical orientation as $\cos2 \theta_c =(1- \kappa)/2(1+\kappa)$, with $\kappa=\sigma_1/\sigma_2$ being the ratio between the major and minor principal stresses on the sample \citep{Griffith1921}. 
For the diametral point load, in which $\sigma _2 = 0$, we can easily deduce that $\theta_c = 30^{\circ}$. 
Nonetheless, in our tests $\theta_c$ varies with $\eta$ from $\simeq 30^{\circ}$ for $\eta = 2$, to $\simeq 15^{\circ}$ for higher values of $\eta$ (see inset of Fig. \ref{fig:crushing_strength}).
These observations show that our numerical experiments are in good agreement with experimental testing and analysis, despite the fact that the `\emph{U}' shape in our results is more subtle than what is reported in literature.
The increasing disagreement in $\theta_c$ with respect to the theoretical estimation also suggests that the inherent anisotropy deeply modifies the stress configuration within the samples. 

For the case without inherent anisotropy, we can follow classical rock mechanics testing and use the maximum tensile stress criterion to characterize the strength of our samples \citep{Hiramatsu1966,Jaeger1967}. 
So the expression $\sigma_c = 2 F_c / \pi d$ allows one to deduce the maximum tensile stress at the center of the circular sample. 
Using that equation, we find that, on average, $\sigma_c/C_n \simeq 1$. 
This result allows us to make two important observations: 1) our model correctly scales the internal material strength ($C_n$) to the macroscale, and 2) it can be correctly assumed that for a non-anisotropic brittle structure, the tensile stresses are indeed triggering the failure. 
However, experimental testing characterizing inherently anisotropic materials should be aware that the tensile stresses are not necessarily at the origin of failure. 

\subsection{Statistical variability of strength}
To better understand the variability of failure strength, we analyze our results in terms of `survival' using the Weibull probability distribution. 
This approach assumes that a sample's probability of not presenting failure $P_s$ (i.e., the survival) depends on the applied stress $\sigma$ as 
\begin{equation}\label{eq:weibull}
P_s = \exp \left\{-\left(\frac{\sigma}{\sigma_0}\right)^m\right\},
\end{equation}

with $\sigma_0$ being a reference stress for which $P_s = 1/e \simeq 37\%$. 
The exponent $m$ is known as the Weibull modulus and is associated with the sharpness of the probability distribution. 
As $m$ increases, so does the slope of the distribution, meaning that the failure strength is focused on a given value. 
Conversely, as $m$ decreases, the stress range within which the particle may break broadens. 

\begin{figure}
    \centering
    \includegraphics[width=0.55\linewidth]{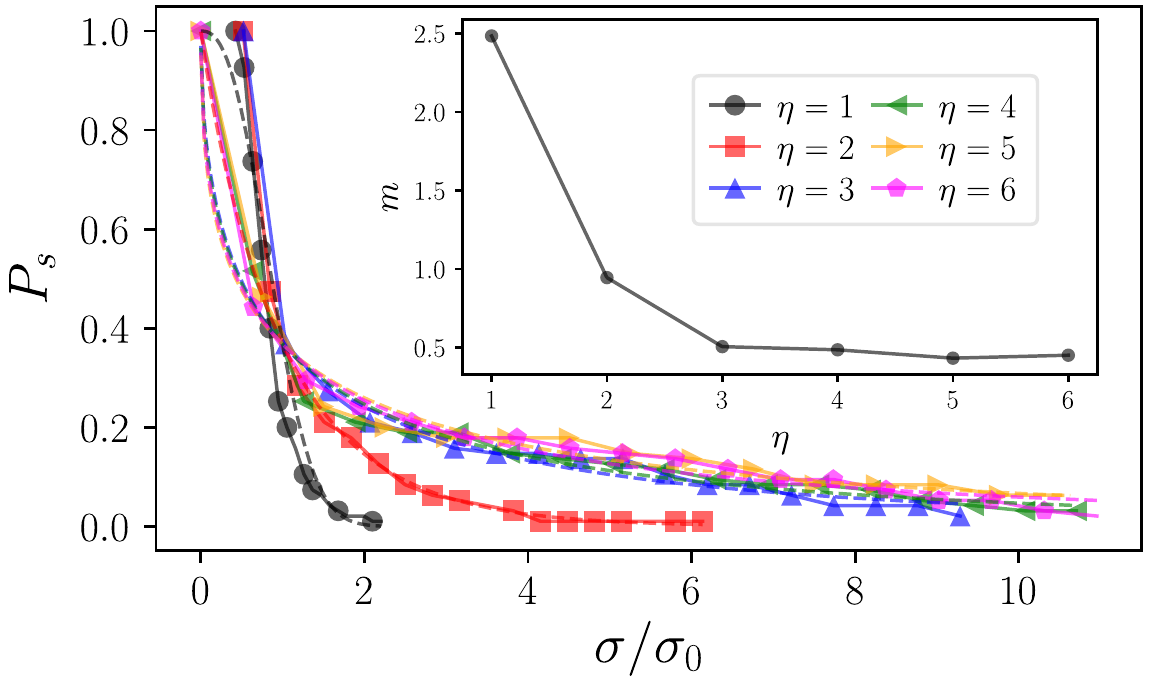}
    \caption{Probability of survival $P_s$ as a function of the applied stress $\sigma$ for each value of $\eta$ and all values of $\theta$ combined. 
    The dashed lines correspond to the fit of Eq. (\ref{eq:weibull}) to the data. 
    In the inset, we present the corresponding value of the Weibull modulus for each fit of the Weibull distribution.} 
    \label{fig:weibull}
\end{figure}

Figure \ref{fig:weibull} presents the survival probability distribution $P_s$ as a function of the applied stress and the different values of $\eta$, combining all results by $\theta$. 
As expected, the stress range $\sigma$ within which we can expect failure considerably increases with $\eta$. 
The dashed lines correspond to the fitting of Eq. (\ref{eq:weibull}) by finding $m$ and $\sigma_0$ with a least-squares minimization. 
The inset on the same figure presents the values found for the Weibull modulus $m$ as a function of $\eta$. 
Typical values for parameter $m$ for silicate materials are found in the range $[1.5,4]$ \citep{McDowell1998,Lim2004}. 
We observe that for microstructures with $\eta=1$, $m$ reaches a value of $\simeq 2.5$ and then smoothly decreases as values of inherent anisotropy grow. 
It is remarkable that our experiments satisfactorily reproduce the Weibull modulus for brittle silicate materials despite the strong variation of the cells' configuration.

If we combine all values of failure strength - not distinguishing between $\eta$ and $\theta$ - we find that $m \simeq 1.5$, which falls within a typical range of values for rocks or grains that are not necessarily anisotropic.
This observation suggests that laboratory tests in which no special attention is given to the degree of inherent anisotropy or loading orientation can gather a wide variety of material characteristics. 
Such a simplified approach could thus be misleading and limit the predictability of the material failure strength. 

\subsection{Macroscopic failure modes and energy consumption}
\begin{figure}
    \centering
    {$\bm{\theta=0^{\circ}}$} \hspace{2.5cm} {$\bm{\theta=30^{\circ}}$} \hspace{2.5cm} {$\bm{\theta=60^{\circ}}$} \hspace{2.6cm} {$\bm{\theta=90^{\circ}}$} \\
    \subfigure {\label{fig:frac00}
    \includegraphics[width=0.22\columnwidth]{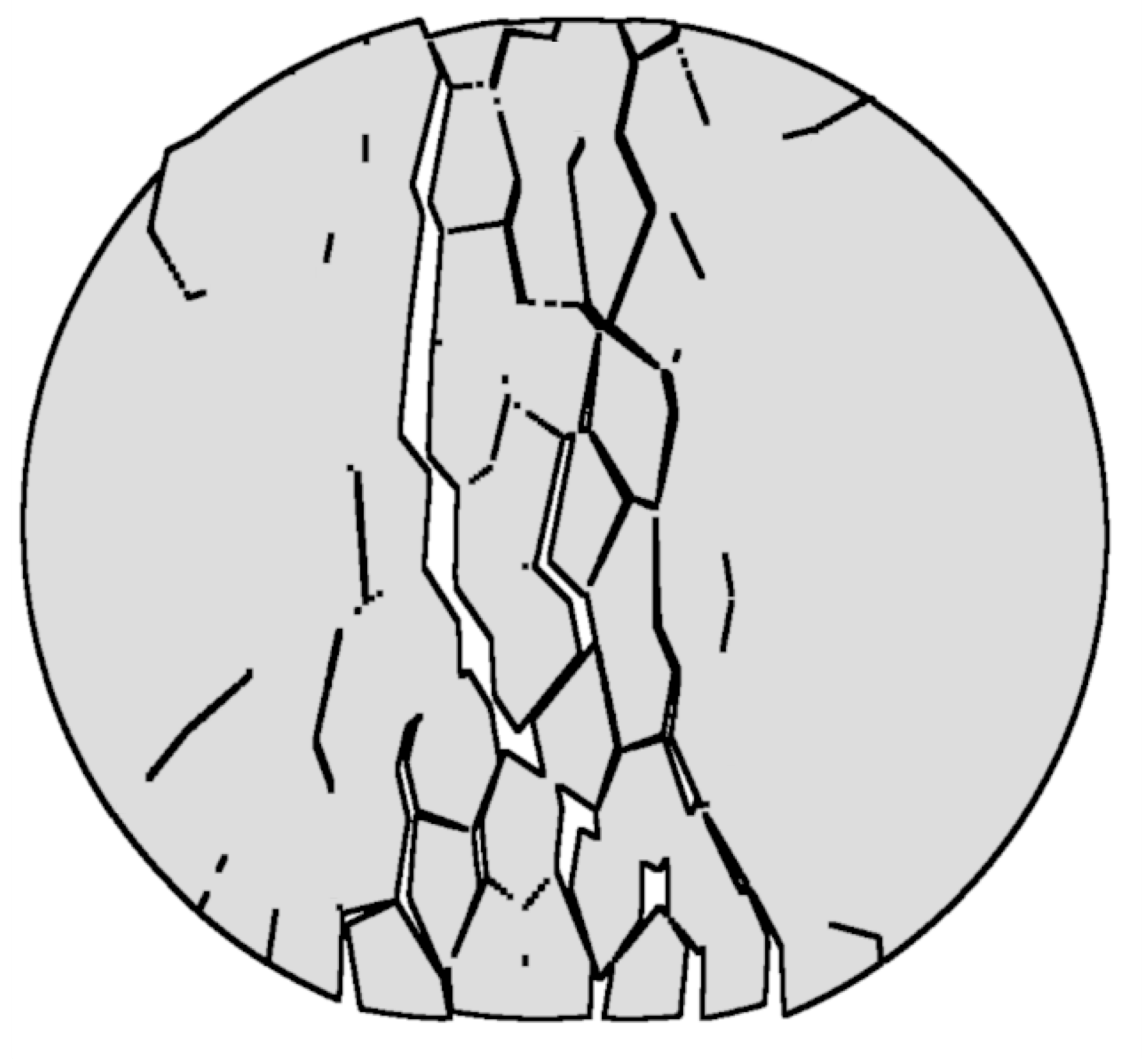}}
    \subfigure {\label{fig:frac30}
    \includegraphics[width=0.22\columnwidth]{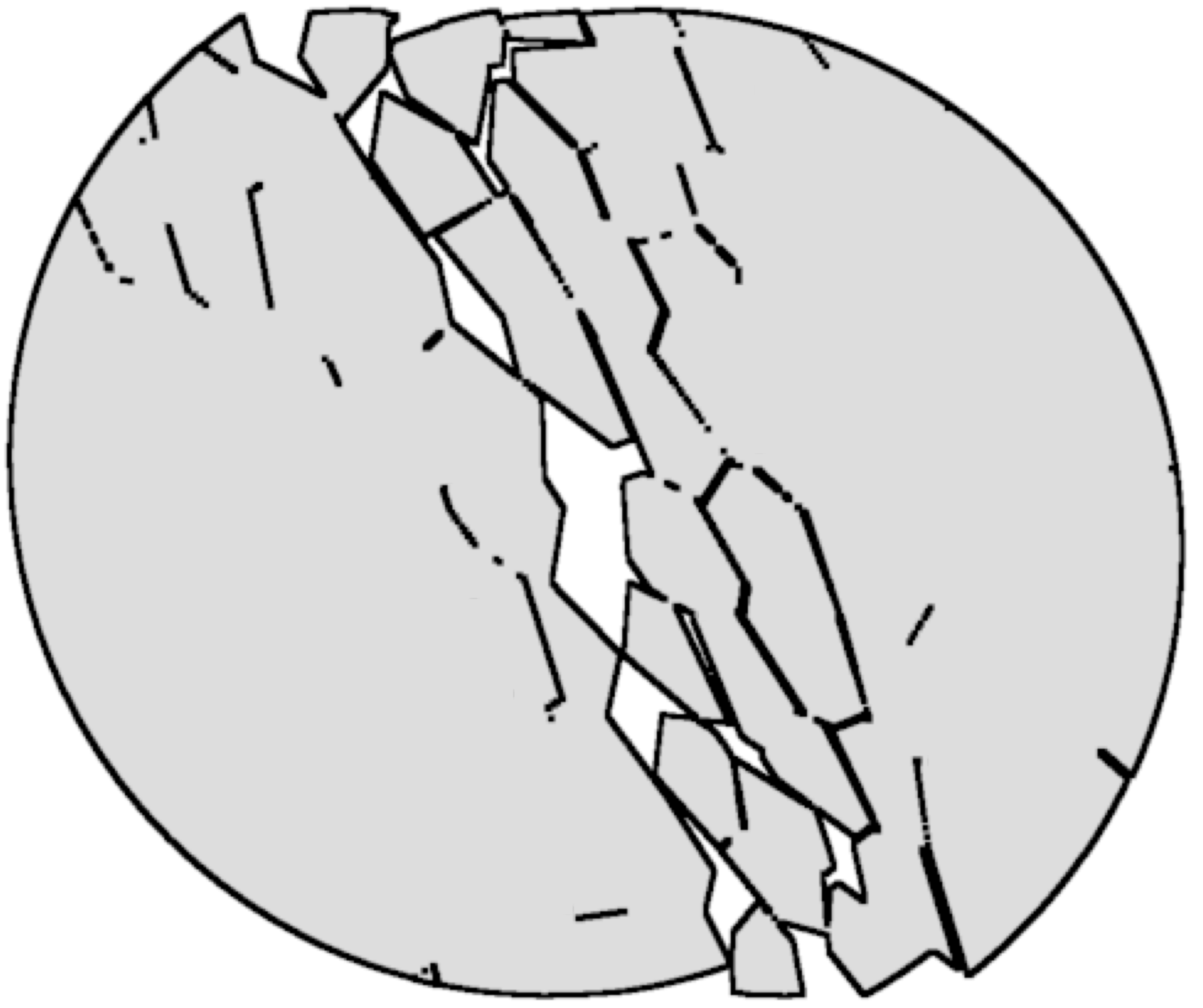}}
    \subfigure {\label{fig:frac60}
    \includegraphics[width=0.21\columnwidth]{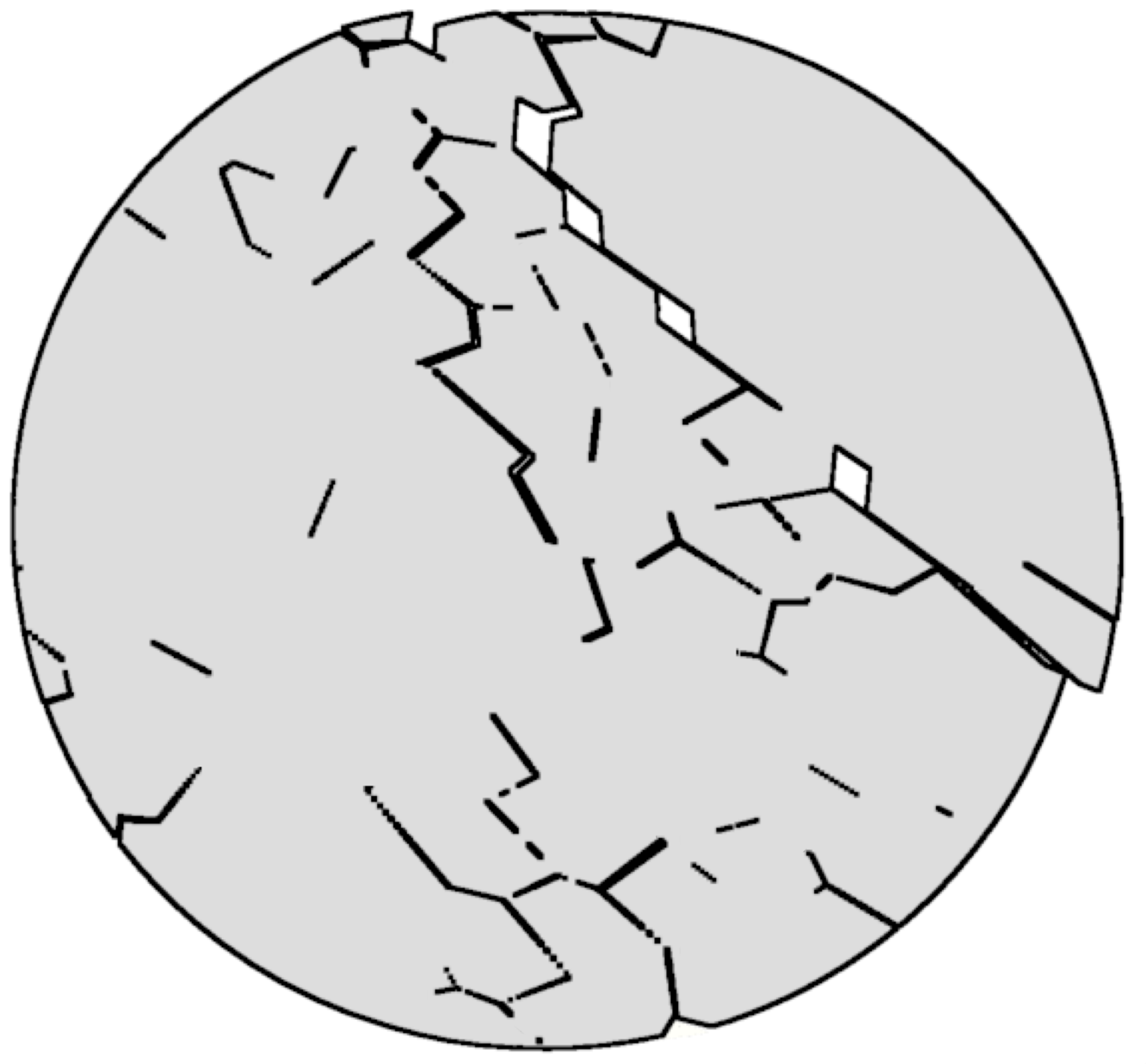}}
    \subfigure {\label{fig:frac90}
    \includegraphics[width=0.21\columnwidth]{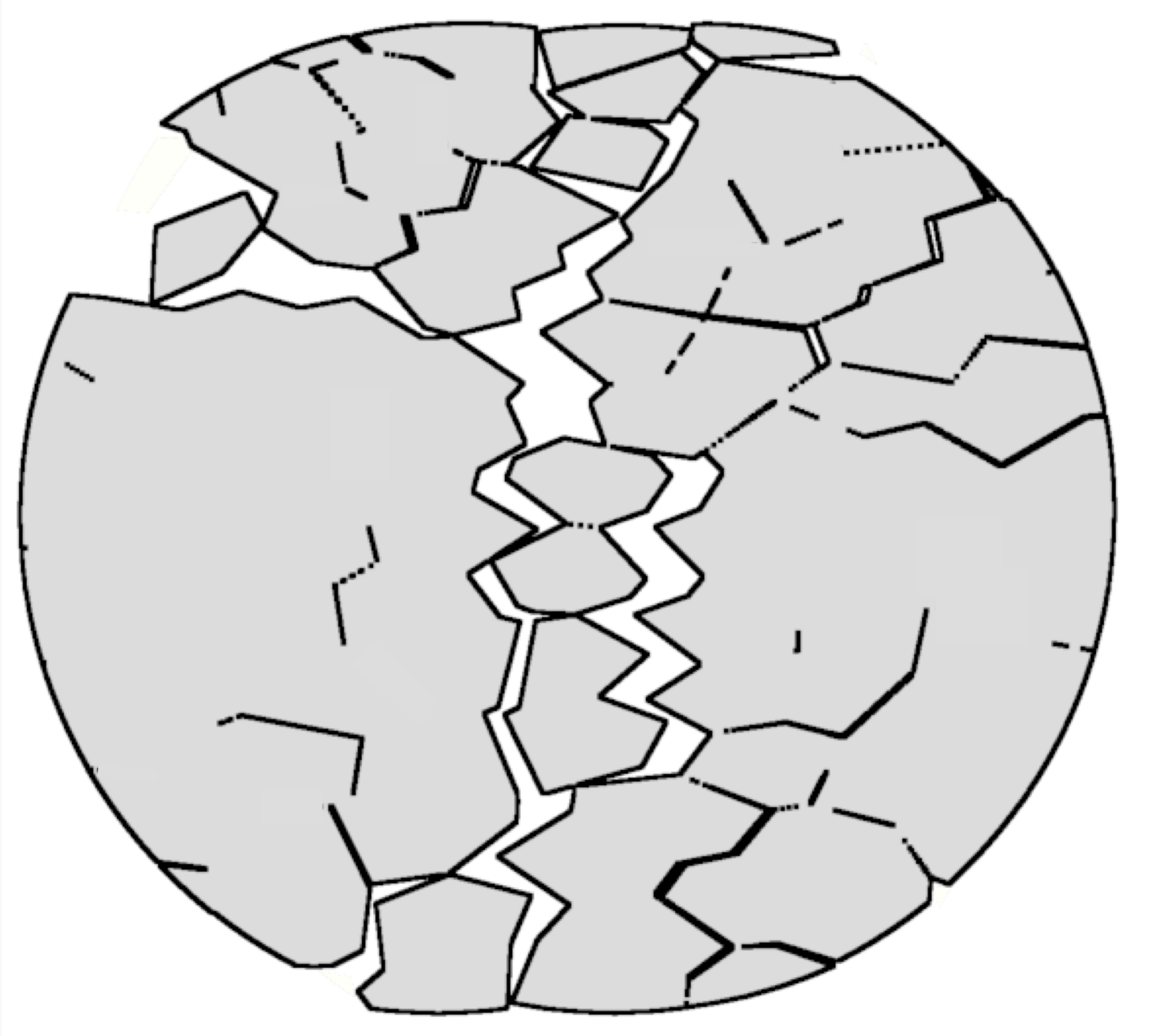}}
    \caption{Evolution of the failure mechanism as the loading orientation increases from $\theta = 0^{\circ}$ up to $\theta = 90^{\circ}$ for the sample with inherent anisotropy $\eta=2$.}
    \label{fig:failure_modes}
\end{figure}

Figure \ref{fig:failure_modes} presents the fissuring paths for some of the samples with $\eta=2$ and varying loading orientation $\theta$. 
When $\theta = 0^{\circ}$, the failure is roughly vertical, matching the loading orientation. 
The failure mechanism is similar when $\theta = 90^{\circ}$, although the zones in contact with the platens show more damage. 
In these two cases, we can infer that tensile stresses are the source of fissuring because many interactions are debonded orthogonally to the loading direction. 
However, when $\theta = 30^{\circ}$ and $60^{\circ}$, the failure mode is different. 
The fissuring is diagonal to the loading, which suggests that shearing is the preferred fissuring mode. 
This observation is in agreement with physical experiments and helps to justify the substantial drop in failure strength $\sigma_{yy}/C_n$ for inherently anisotropic structures. 

Note that this observation about the macroscopic failure modes is descriptive. 
A quantitative assessment of the energy consumption shows that all the samples consume roughly the same amounts of fragmentation energy independently of $\theta$ and $\eta$ \citep{Cantor2021}. 
In fact, the similarity of the total length of cohesive bonds is the central parameter controlling the energy consumption in the samples. 
Then, the variability in failure strength must be linked to microstructural elements rather than to the mechanisms splitting failures between tensile and shearing modes. 

\section{Microstructural analysis}\label{Sec:Micro}
As previously mentioned, rock microstructure is often related to mineral or graining size and shape distribution, joint spacing/density, fissuring, and etc. 
All of these geometrical characteristics clearly affect the failure strength. 
However, the microstructure cannot only be reduced to its geometrical aspects. 
Accounting for the connectivity between cells and the force transmission are key elements behind the macroscopic mechanical behavior.

In order to do this, we first need to define a framework of analysis. 
We have two possibilities when dealing with adjacent cells. 
First, the \emph{interaction} frame in which the bond forces are defined as $\bm{f} = f_n \bm{n} + f_t \bm{t}$, with $\bm{n}$ being the normal unit vector perpendicular to the contact line, and $\bm{t}$ being the tangential unit vector. 

\begin{figure}
    \centering
    \includegraphics[width=0.42\columnwidth]{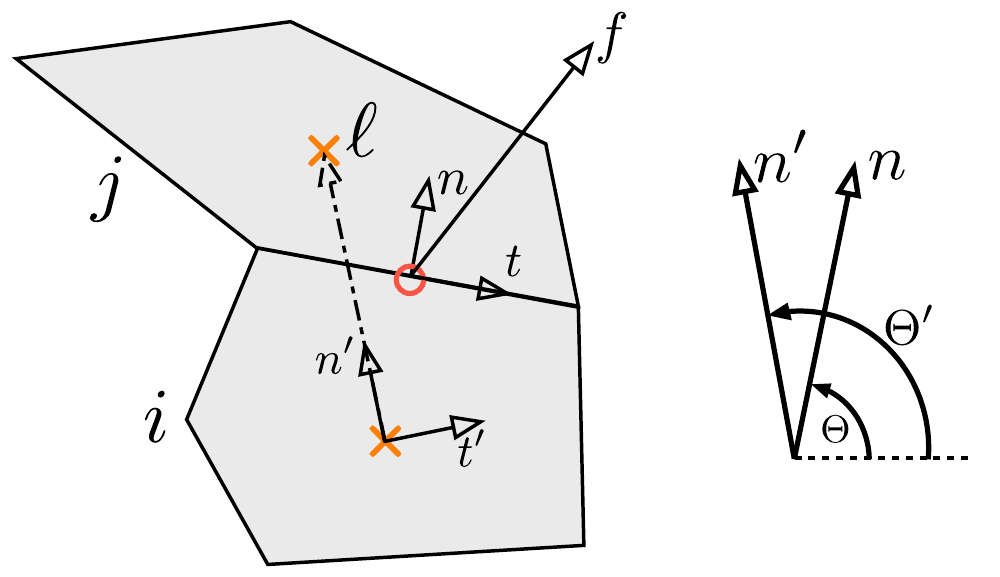}
    \caption{Schematic representation of the local frames created between two irregular cells $i$ and $j$. Unitary vectors $\bm{n}$ and $\bm{t}$ are linked to the bond, while $\bm{n^{\prime}}$ and $\bm{t^{\prime}}$ are defined upon the branch vector. Note that angles $\Theta$ and $\Theta^{\prime}$ are the orientation of $\bm{n}$ and $\bm{n^{\prime}}$, respectively, measured counterclockwise from the horizontal.}
    \label{fig:ctc_branch}
\end{figure}

We can also define the inter-center vectors between cells, also called \emph{branch vectors}, as $\bm{\ell} = \ell_n \bm{n} + \ell_t \bm{t}$, with $\ell_n$ and $\ell_t$ being the normal and tangential components. 
These branches let us define a second frame in which the unit vector $\bm{n^{\prime}}$ is defined along $\bm{\ell}$, and $t^{\prime}$ is the tangential unit vector (see Fig. \ref{fig:ctc_branch}) \citep{Azema2010}. 
In this frame, the forces between the cells are written as $\bm{f} = f^{\prime}_n \bm{n^{\prime}} + f^{\prime}_t \bm{t^{\prime}}$, with $f^{\prime}_n$ and $f^{\prime}_t$ - the radial and ortho-radial forces, respectively - acting between the centers of the cells. 
Finally, the branch in this frame is simply written as $\bm{\ell} = \ell \bm{n^{\prime}}$, with $\ell$ being the length of the branch vector. 
For convenience, we used the branch frame for the following microstructural analysis. 

\begin{figure}
    \centering
    {$\bm{\eta=1}$} \hspace{3.2cm} {$\bm{\eta=3}$} \hspace{3.2cm} {$\bm{\eta=6}$} \\
    \raisebox{0.65in}{\rotatebox[origin=t]{90}{$\bm{\mathrm{Bonds \ Network}}$}}
    \subfigure{
    \includegraphics[width=0.18\columnwidth]{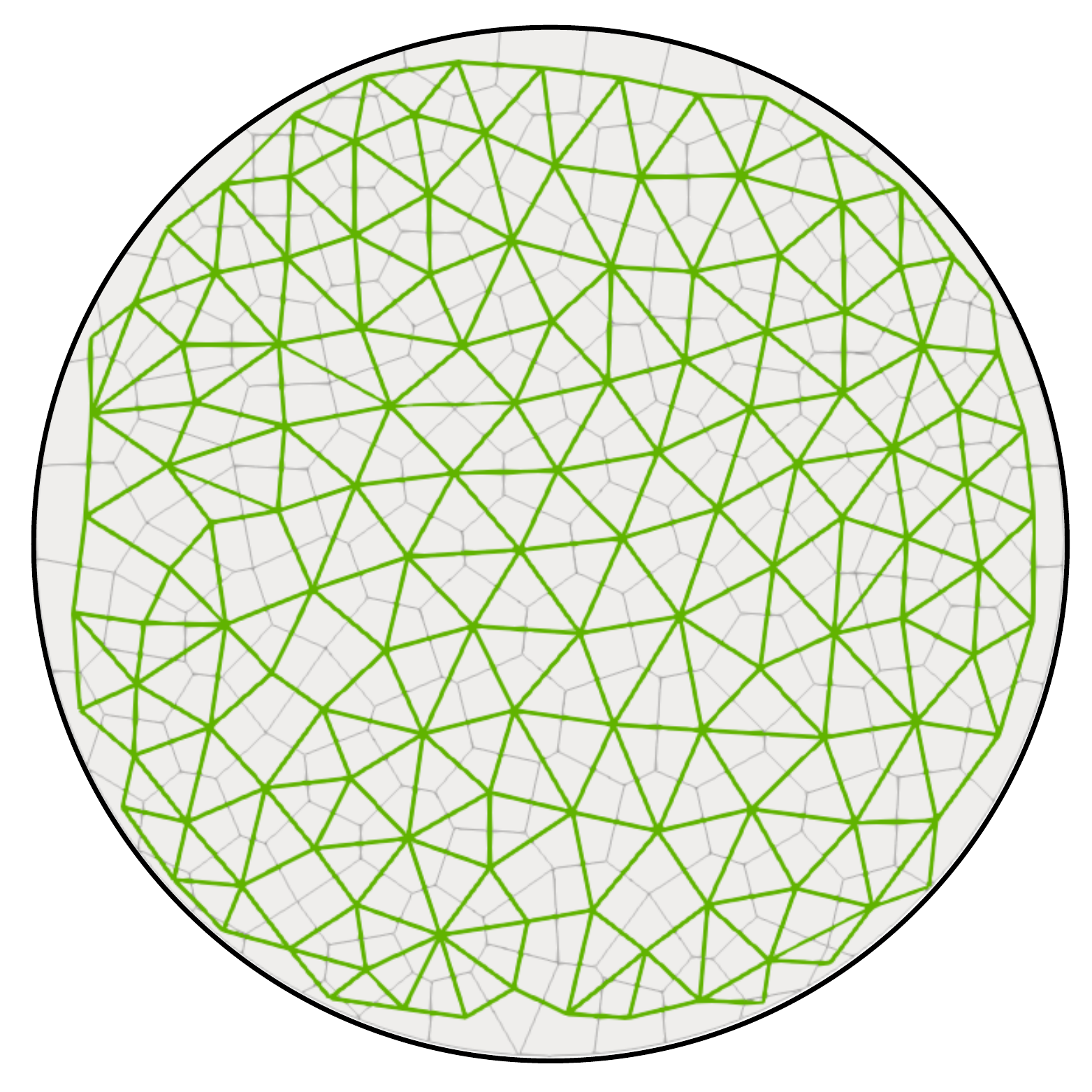}}
    \hspace{0.48cm}
    \subfigure{
    \includegraphics[width=0.18\columnwidth]{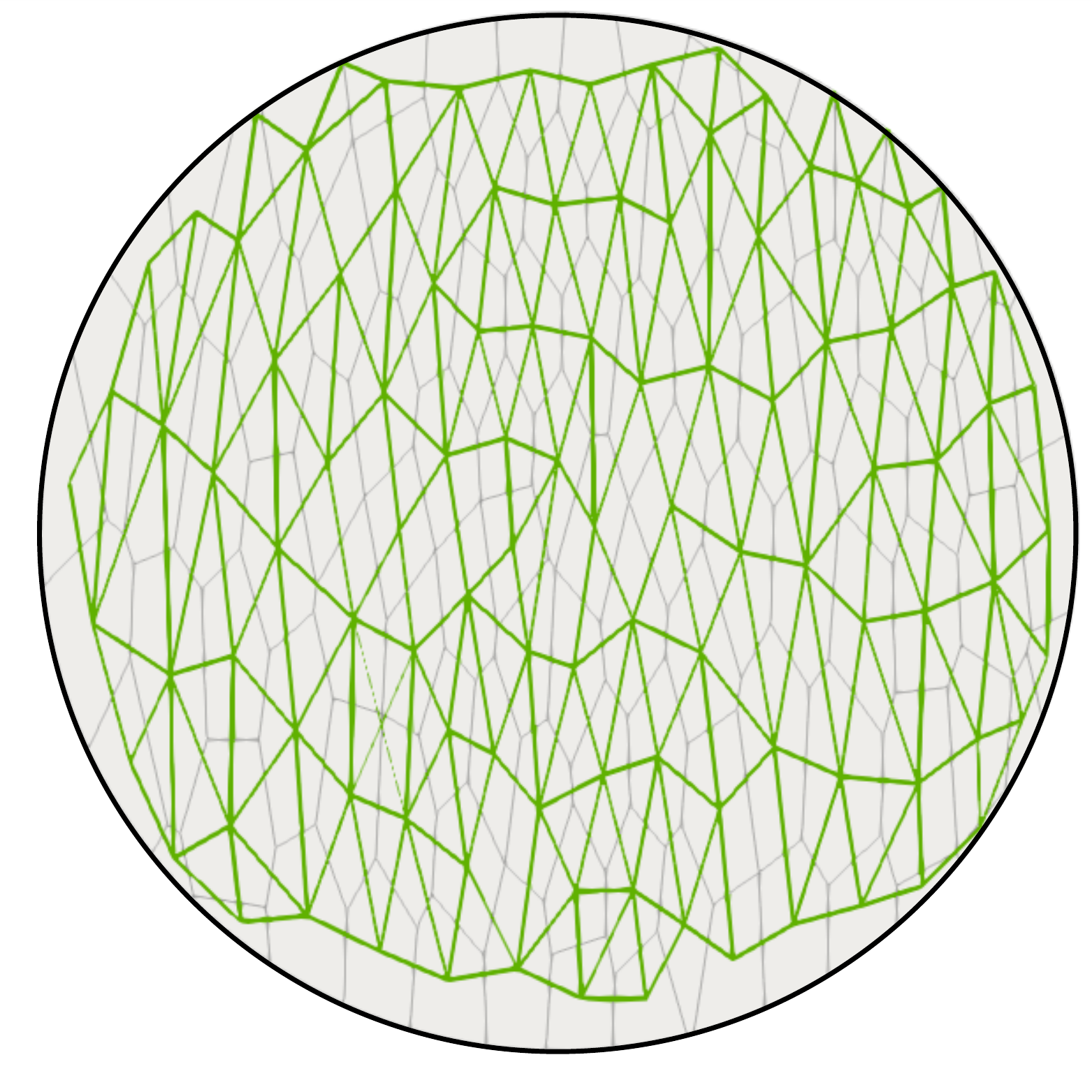}}
    \hspace{0.48cm}
    \subfigure{
    \includegraphics[width=0.18\columnwidth]{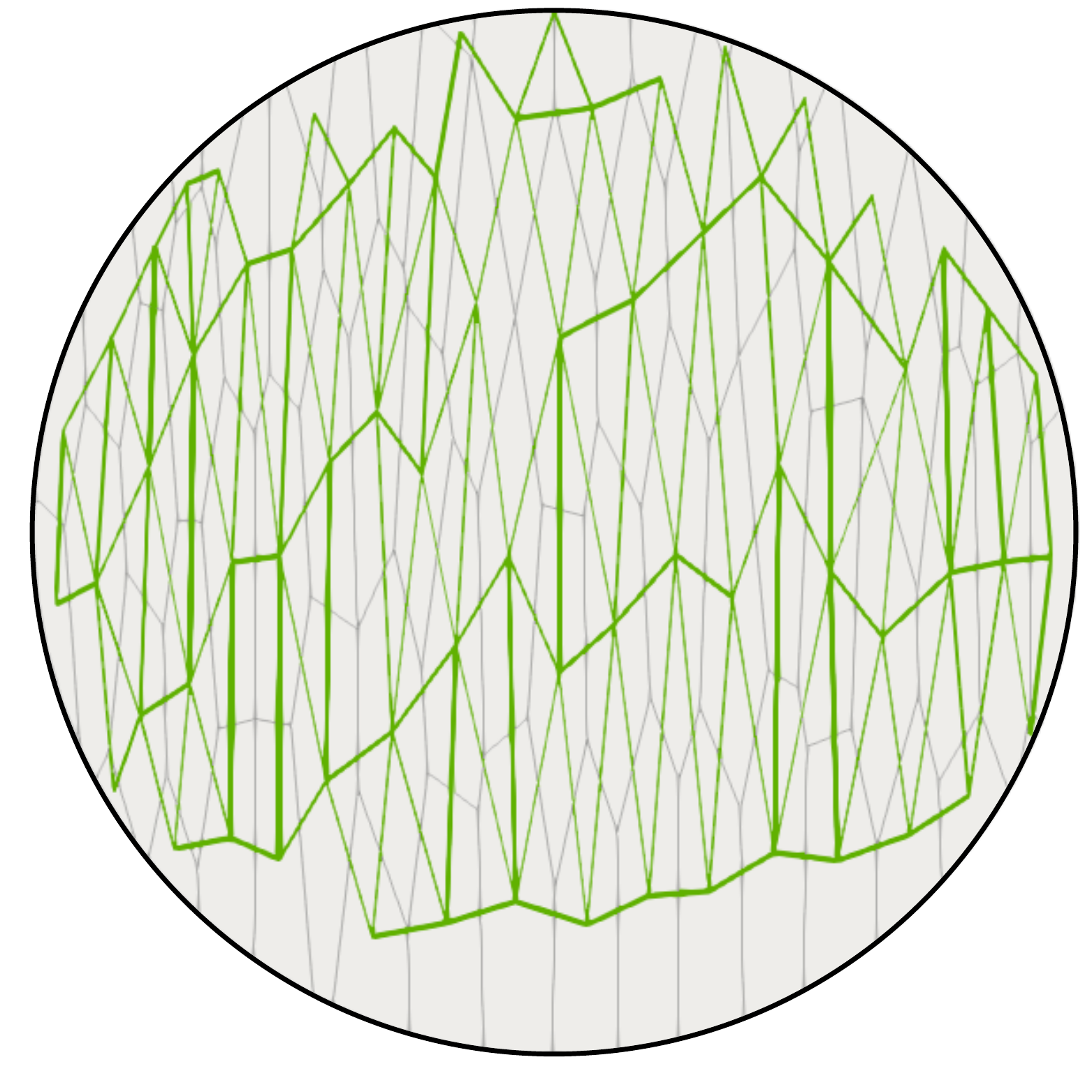}}
    \\
    \raisebox{0.65in}{\rotatebox[origin=t]{90}{$\bm{\mathrm{Forces \ Network}}$}}
    \subfigure{
    \includegraphics[width=0.18\columnwidth]{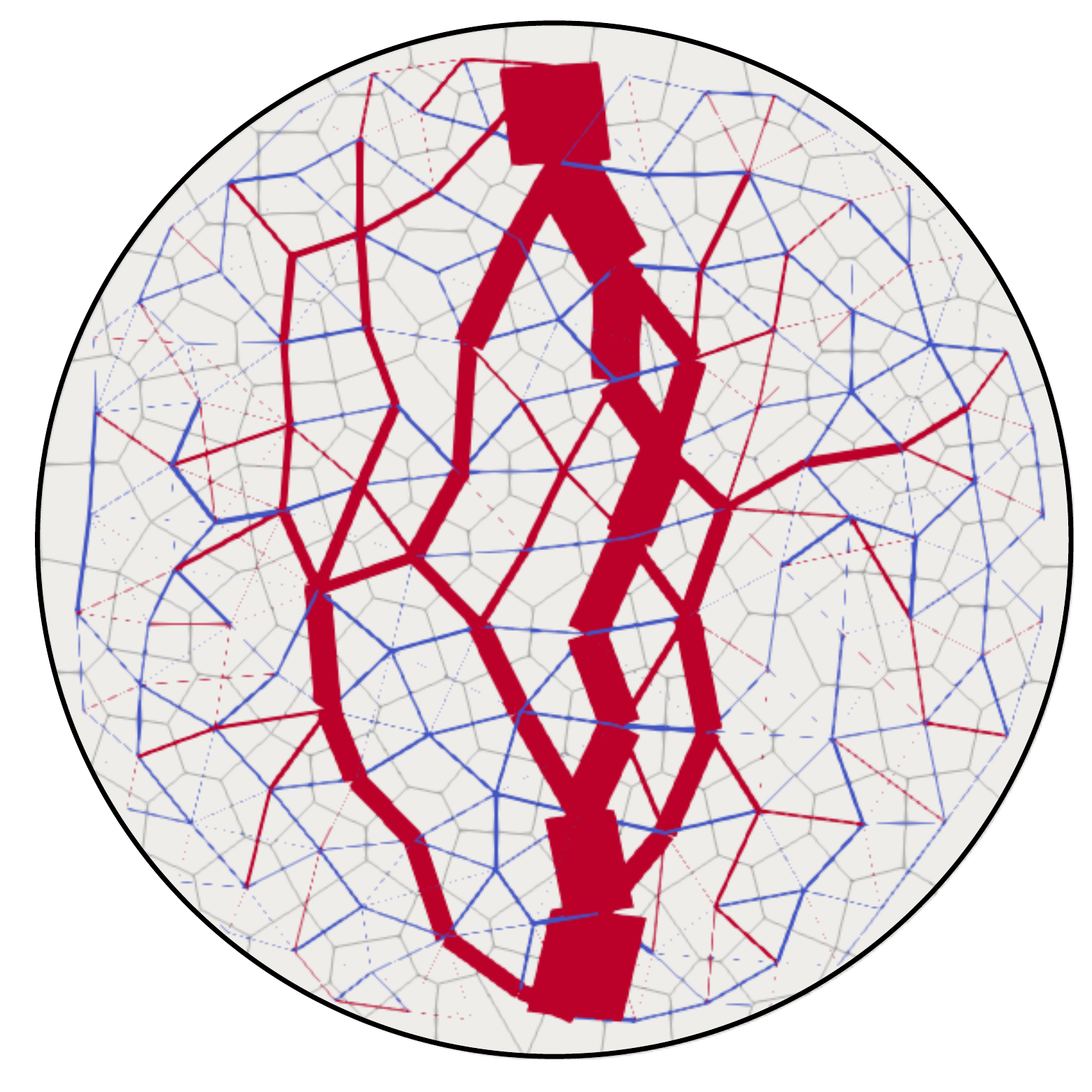}}
    \hspace{0.48cm}
    \subfigure{
    \includegraphics[width=0.18\columnwidth]{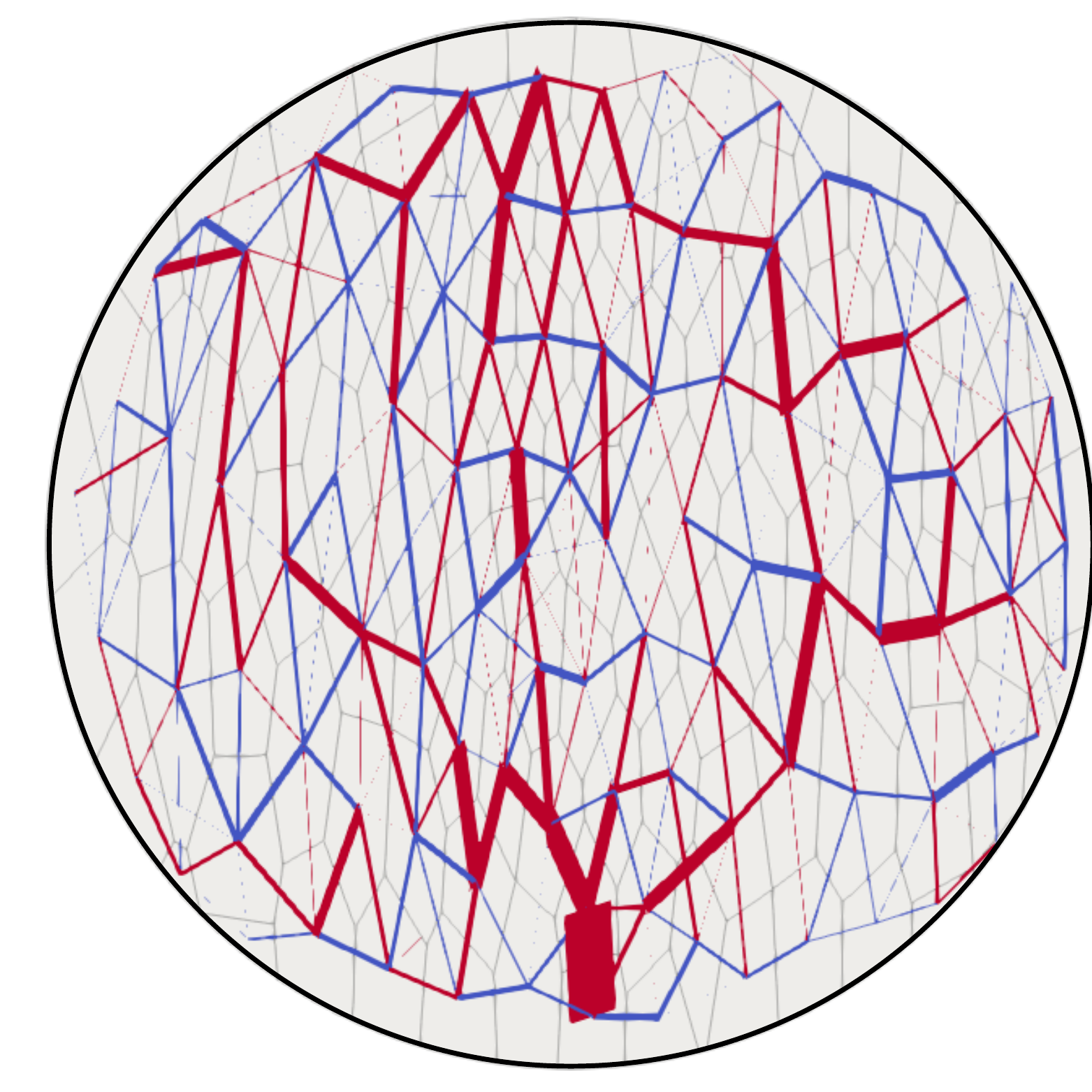}}
    \hspace{0.48cm}
    \subfigure{
    \includegraphics[width=0.18\columnwidth]{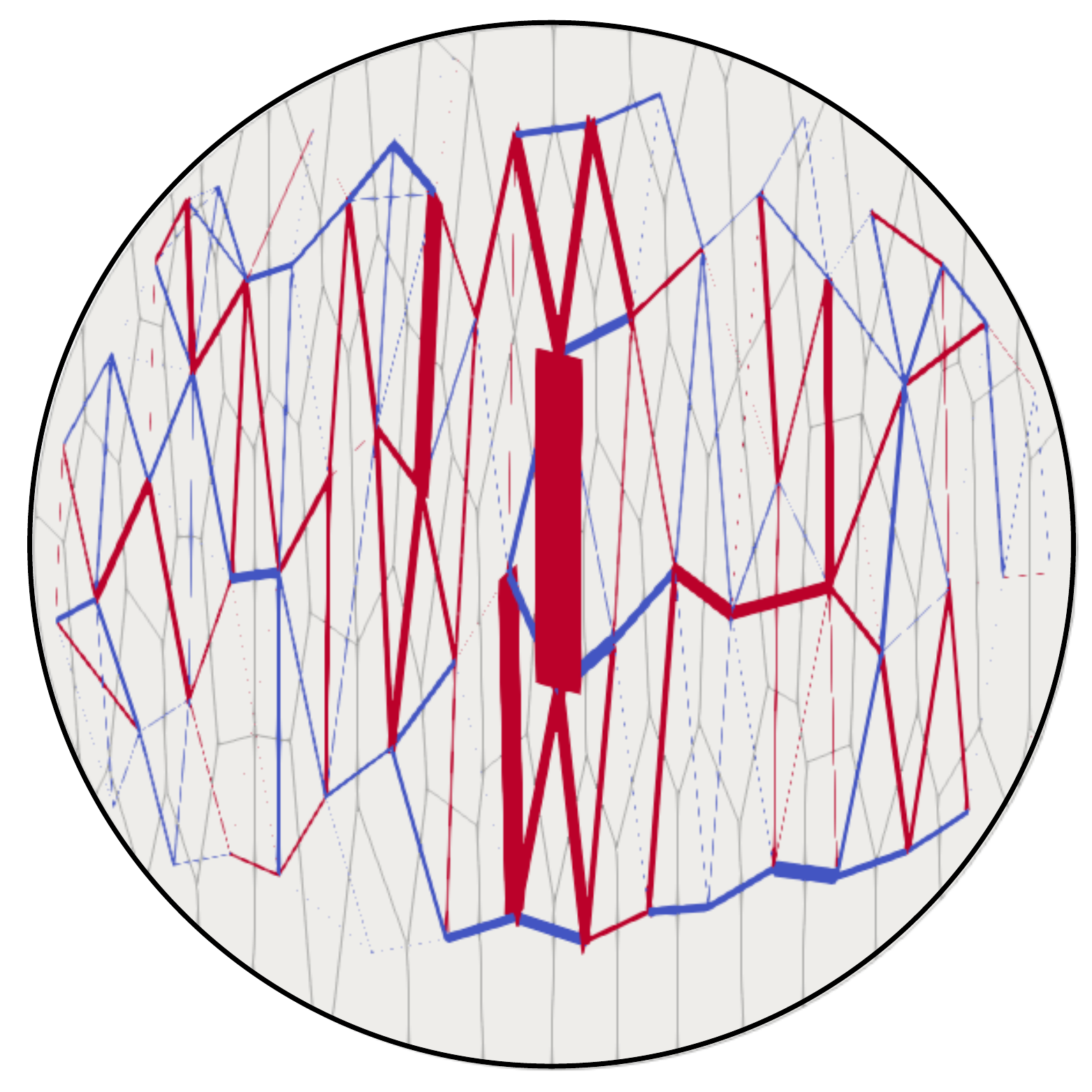}}
    \caption{(Top) Networks created by bonded adjacent cells for $\eta=1,3$ and $6$ displayed with lines between the center of mass of the corresponding cells. (Bottom) Force networks between bonded cells. The thickness of the lines is proportional to the force intensity. Traction forces are displayed in blue and compression forces in red. These screenshots are taken for cases in which the loading orientation is $\theta=0$.}
    \label{fig:micro_ex}
\end{figure}

As an illustration, Fig. \ref{fig:micro_ex} (top) presents the branch network and the force network (bottom) with lines whose thickness is proportional to the intensity of the force at the interactions. 
A visual inspection shows how the geometry of the cells dramatically modifies both networks. 
As $\eta$ increases, the branch network becomes more irregular and the force chains more diffuse within the volume. 

\subsection{Geometrical description}
\subsubsection{Connectivity}
We can characterize the connectivity between cells by using the coordination number $Z$. 
This parameter shows the average number of neighboring interactions per cell as $Z = 2 N_c/N_{cl}$, with $N_c$ being the number of bonds between cells and $N_{cl}$ the number of cells. 
There is, however, a subtle difference in coordination number between the intact state (i.e., at the beginning of the loading) and the coordination number we are able to compute instants before failure. 
Both values differ, as fissuring removes cohesive bonds previously capable of bearing force. 
We characterize the onset of failure as the state bearing $\sigma_{yy}$, so let us consider the cohesive bonds at the onset of failure $N^*_c$ as the effective number of interactions, so $Z^* = 2 N_c^*/N_{cl}$. 

\begin{figure}
    \centering
    \includegraphics[width=0.5\columnwidth]{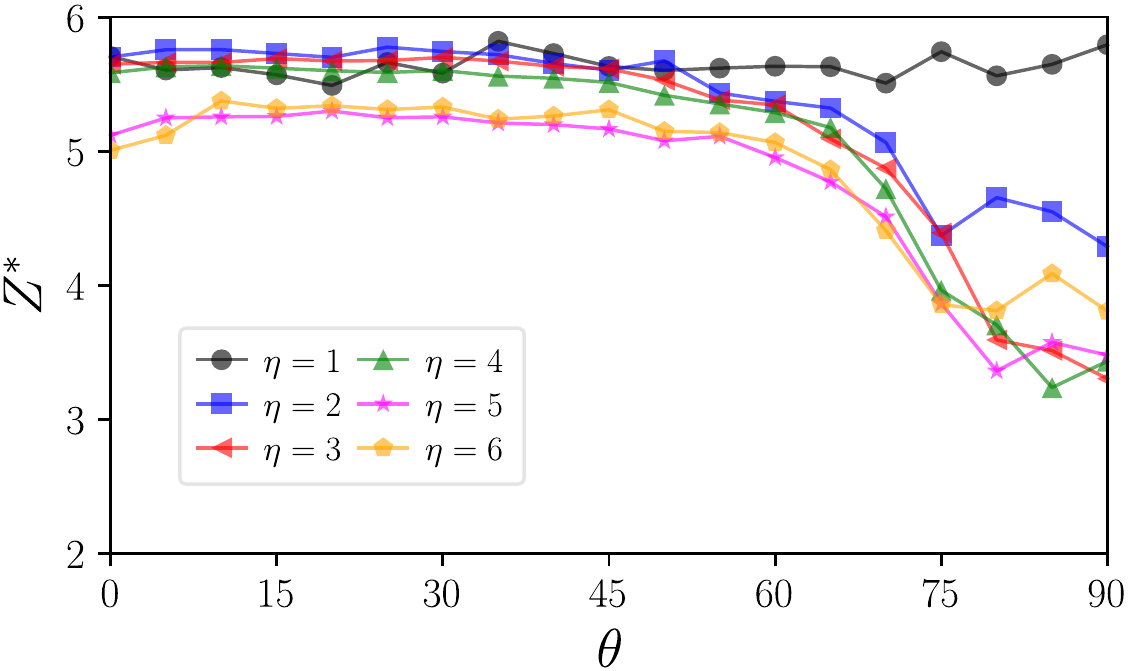}
    \caption{Evolution of the coordination number at the onset of failure $Z^*$ as a function of the loading orientation $\theta$ and the inherent anisotropy $\eta$.}
    \label{fig:z}
\end{figure}

Figure \ref{fig:z} presents the averaged coordination number at the onset of failure as a function of $\eta$ and $\theta$. 
We observe that the connectivity decreases as $\theta$ and $\eta$ increase. 
In other words, $Z^*$ varies conversely to $\sigma_{yy}$ showing that inherently anisotropic materials can bear larger stresses despite the fact that cells are less connected. 
The evolution of $Z^{*}$ provides a counterintuitive picture of the effect of $\theta$ and $\eta$ on the microstructural properties at failure. 
However, as previously shown in Fig. \ref{fig:ctc_branch}, the branch and force networks also carry a strong anisotropic character that calls for a higher-order analysis accounting for their distribution in space.

\subsubsection{Branch orientations}
\begin{figure}
    \centering
    \subfigure {\label{fig:ac}(a)
    \includegraphics[width=0.45\columnwidth]{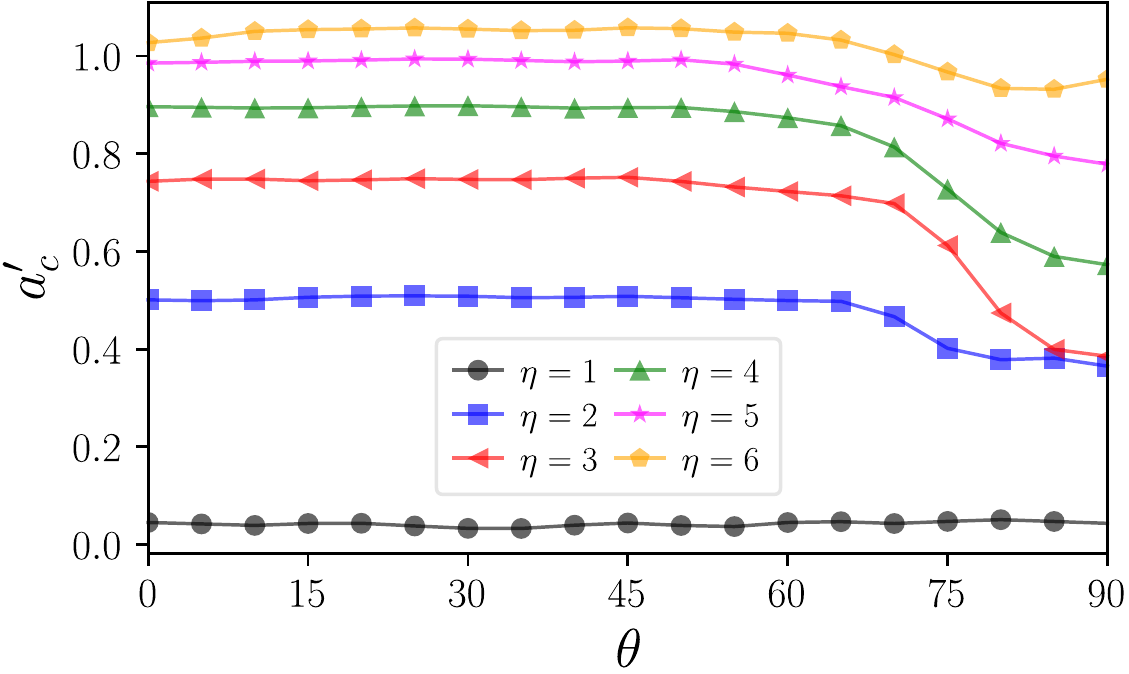}}
    \subfigure {\label{fig:t_ac}(b)
    \includegraphics[width=0.45\columnwidth]{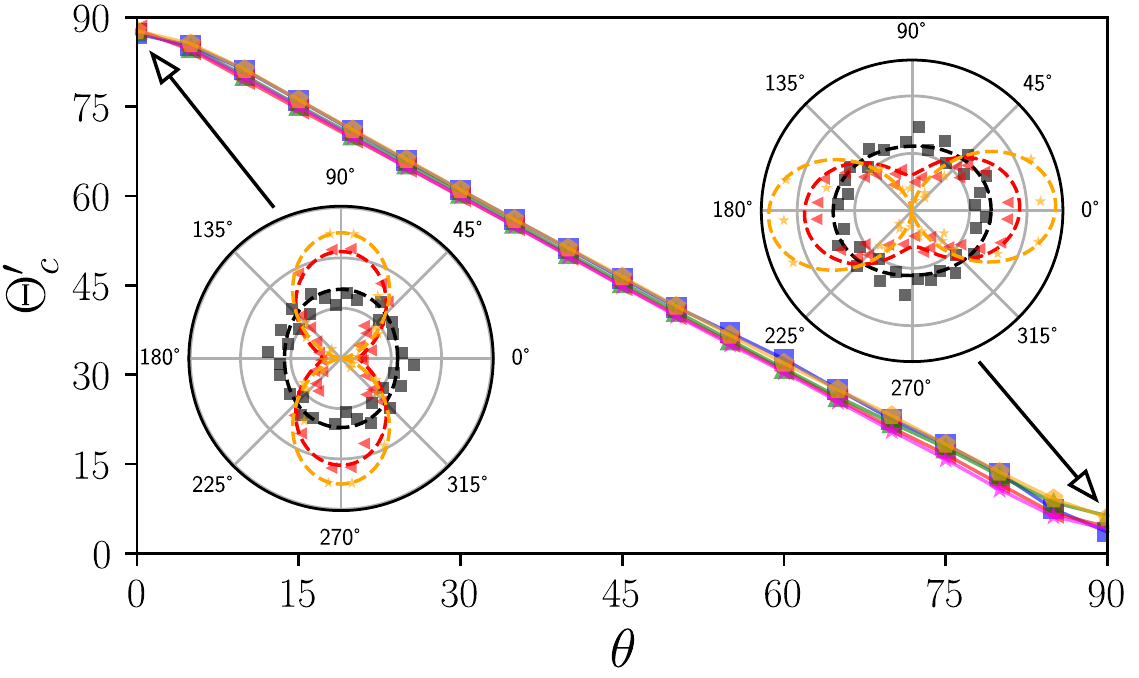}}
    \caption{Bond orientation anisotropy $a^{\prime}_c$ (a) and the preferential orientation $\Theta^{\prime}_c$ (b) for different inherent anisotropy levels $\eta$ and loading orientations $\theta$.
    In the inset: probability of branch orientations $P_c(\Theta^{\prime})$ shown with symbols for three values of inherent anisotropy $\eta$ and loading orientation $\theta = 0^{\circ}$ and $\theta = 90^{\circ}$. The dashed lines are the fitting curves using Eq. (\ref{eq:P_c}).}
    \label{fig:ctc_aniso}
\end{figure}

We can define the density probability distribution $P_c$ of branch vector orientations as 
\begin{equation}
P_c(\Theta^{\prime}) = \frac{N^*_{c}(\Theta^{\prime})}{N^*_{c}}
\end{equation}

with $N^*_{c}(\Theta^{\prime})$ being the number of branches pointing at angle $\Theta^{\prime}$ at the onset of failure. 
The inset of Fig. \ref{fig:t_ac} presents the angular distributions $P_c(\Theta^{\prime})$ with symbols for three different values of inherent anisotropy and loading orientations $\theta = 0^{\circ}$ and $\theta = 90^{\circ}$. 
We can see that when $\eta = 1$ the distribution remains almost circular, highlighting the fact that the bond network is nearly isotropic and independent of the assembly's rotation. 
Conversely, the distributions for $\eta >1$ present preferential orientations matching the preferred orientation of the cells. 

These angular distributions can also be described using truncated Fourier series, as 
\begin{equation}\label{eq:P_c}
P_c(\Theta^{\prime}) = \frac{1}{2\pi} \left\{1 + a^{\prime}_c \cos2 \left(\Theta^{\prime}-\Theta^{\prime}_c \right) \right\},
\end{equation}

with $\Theta^{\prime}_c$ being the preferential orientation of the distribution and $a^{\prime}_c$ its anisotropy level, i.e., the branch vector orientation anisotropy. 
Note that $a^{\prime}_c = 0$ means a circular distribution $P_c$, in which bonds are equally presented in all orientations $\Theta^{\prime}$. 
Conversely, as $a^{\prime}_c$ increases, more bonds present a preferential orientation in space. 
Although we could fit Eq. (\ref{eq:P_c}) to our measures to find $a^{\prime}_c$ and $\Theta^{\prime}_c$, we prefer to use the \emph{fabric} tensor defined as \citep{Rothenburg1989}
\begin{equation}\label{eq:fabric}
\bm{F}_{ij} = \int_0^{\pi} P_c(\Theta^{\prime}) n^{\prime}_i(\Theta^{\prime}) n^{\prime}_j(\Theta^{\prime}) d\Theta^{\prime},
\end{equation}

with $\bm{n^{\prime}} = \{\cos\Theta^{\prime}, \sin{\Theta^{\prime}}\}$.
Equation (\ref{eq:fabric}) lets us define the anisotropy of branch orientations as $a^{\prime}_c = 2(F_1 - F_2)$, with $F_1>F_2$ being the eigenvalues of $F$. 
The major principal direction of the fabric tensor is $\Theta^{\prime}_c = 1/2 \arctan \{2 F_{xy}/(F_{xx} - F_{yy})\}$, with $F_{xx}$ and $F_{yy}$ being the components in the diagonal of $\bm{F}$, and $F_{xy}$ the component off the diagonal. 
Figure \ref{fig:ctc_aniso} presents the corresponding values of preferential branch orientation and anisotropy as a function of $\eta$ and $\theta$. 

We can observe that $a^{\prime}_c$ is close to zero for $\eta=1$, exposing the isotropic character of the branch network and its independence before $\theta$. 
For $\eta>1$, $a^{\prime}_c$ can reach larger values as high as $\simeq 1.1$ for $\eta=6$. 
There is also a slight drop in $a^{\prime}_c$ occurring for values of $\theta>70^{\circ}$. 
This phenomenon, combined with the drop of $Z^*$ we observed before, suggests that contacts are mostly lost in the minor orientation of the fabric tensor as $\eta$ increases. 

In Fig. \ref{fig:t_ac}, we present the evolution of $\Theta^{\prime}_c$ as a function of the loading orientation $\theta$. 
For $\eta=1$, $\Theta^{\prime}_c$ is irrelevant given that $a^{\prime}_c \simeq 0$, so it is omitted from the plot. 
But as soon as $\eta>1$, $\Theta^{\prime}_c$ decreases as $\pi/2-\theta$ which matches the cells' orientation. 

\begin{figure}
    \centering
    \subfigure{\label{fig:aln}(a)
    \includegraphics[width=0.45\columnwidth]{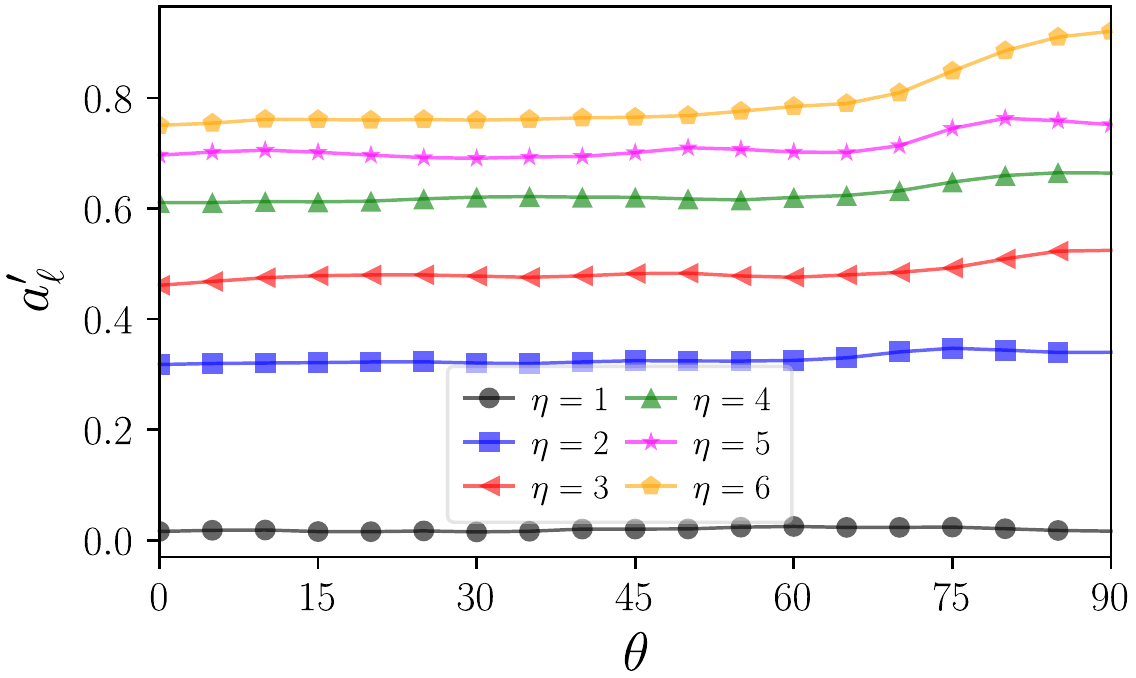}}
    \subfigure{\label{fig:t_aln}(b)
    \includegraphics[width=0.45\columnwidth]{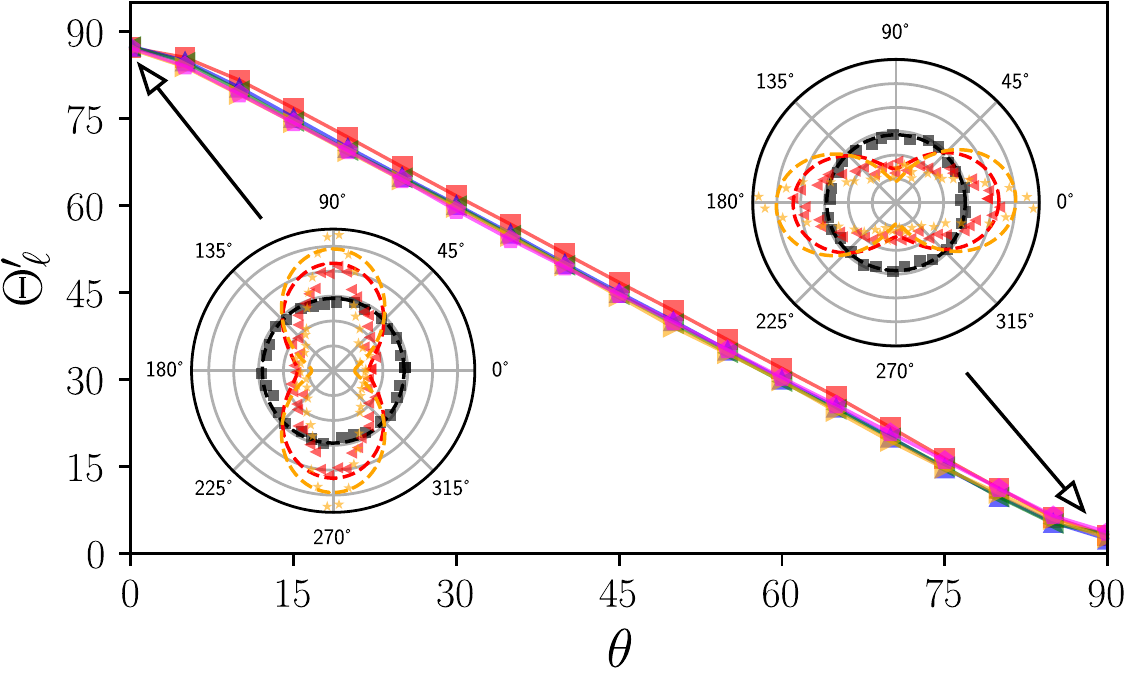}}
    \caption{Evolution of the normal branch length anisotropy (a) and its preferential angular orientation (b) as a function of the loading orientation and values of inherent anisotropy $\eta$. 
    In the inset: angular distribution branch lengths for some values of inherent anisotropy $\eta$ for loading orientations $0^{\circ}$ and $90^{\circ}$. We also present Eq. (\ref{eq:ell_n}), fitting the angular branch length distributions with dashed lines.}
    \label{fig:branch_aniso}
\end{figure}

Along with $P_c$, we can also characterize the angular branch length distribution $\langle \ell \rangle(\Theta^{\prime})$.
This distribution can be computed as a function of $\Theta^{\prime}$ as
\begin{equation}\label{eq:ln_dist}
\langle \ell \rangle(\Theta^{\prime}) = \frac{1}{N^*_{c}(\Theta^{\prime})} \sum_{c \in \delta \Theta^{\prime}} \ell^{c},
\end{equation}

with $\ell_c$ being the length of the branches pointing at small intervals of angular orientation $\delta \Theta^{\prime}$.
The inset in Fig. \ref{fig:t_aln} presents these angular distributions for loading orientations $\theta=0^{\circ}$ and $\theta=90^{\circ}$, and three values of $\eta$. 
These angular distributions for branch lengths closely follow the trends previously seen for $P_c(\Theta^{\prime})$. 

As with branch orientations, the angular evolution of branch lengths can be described using the expression
\begin{equation}\label{eq:ell_n}
\langle \ell \rangle (\Theta^{\prime}) = \langle \ell \rangle \left\{1 + a^{\prime}_{\ell} \cos2 \left(\Theta^{\prime}-\Theta^{\prime}_{\ell}\right) \right\},
\end{equation}
with $\Theta^{\prime}_{\ell}$ being the preferential orientation, and $a^{\prime}_{\ell}$ the level of anisotropy. 
In the insets of Fig. \ref{fig:t_aln}, we show that the branch length distributions become more anisotropic as $\eta$ increases, and that the longest branches predominately point in the same direction as the cells are pointing. 

To find the branch length anisotropy, it is convenient to build the \emph{branch tensor}, which is defined in an integral form as \citep{Rothenburg1989,Azema2011}
\begin{equation}
H^{\ell}_{ij} = \int_0^{\pi} \langle \ell \rangle(\Theta^{\prime}) n^{\prime}_i(\Theta^{\prime}) n^{\prime}_j(\Theta^{\prime}) d\Theta^{\prime}.
\end{equation}

Note that this integral is computed in the range $[0, \pi]$ given the periodic evolution of the angular distributions in that interval. 
We can then compute the branch length anisotropy as $a^{\prime}_{\ell} = 2(H^{\ell}_1 - H^{\ell}_2)/(H^{\ell}_1 + H^{\ell}_2)$, with $H^{\ell}_1$ and $H^{\ell}_2$ being the eigenvalues of the tensor, so $H^{\ell}_1 > H^{\ell}_2$. 
The same construction allows us to compute $\Theta^{\prime}_{\ell}$ as the major principal orientation of $H^{\ell}$ using the same approach as with the fabric tensor. 
In the inset of Fig. \ref{fig:t_aln}, we present Eq. (\ref{eq:ell_n}) with dashed lines using the values extracted from the tensors above, and nicely fitting the angular distributions. 
We deliberately omitted the evolution of $\langle \ell \rangle(\Theta^{\prime})$ for $\eta=1$ since the corresponding values for $a^{\prime}_{\ell}$ are negligible. 

Figures \ref{fig:branch_aniso}(a) and \ref{fig:branch_aniso}(b) gather the results for the branch anisotropies $a^{\prime}_{\ell}$ and preferred orientations $\Theta^{\prime}_{\ell}$ as a function of the loading orientation $\theta$ and the levels of inherent anisotropy $\eta$. 
They show that the branch length anisotropy increases with $\eta$ from $a^{\prime}_{\ell}\simeq 0$ for $\eta=1$, up to $a^{\prime}_{\ell}\simeq 0.7$ for $\eta=6$. 
In all of the cases, these anisotropies present only minor variations with the loading orientation $\theta$. 
We can then say that the variation of the preferred orientation for branch lengths evolves roughly as $\pi/2 - \theta$, similarly to $P_c(\Theta^{\prime})$.

The large variations of geometrical anisotropies that we observed are induced by the inherent anisotropy of the cells - and, furthermore, $\eta$ is likely also inducing force transmission heterogeneities within the samples.
In order to investigate this, we focus next on the interaction forces between cells. 

\subsubsection{Force orientations}
Similarly to the analysis undertaken for branches, we can analyze the interaction forces between cells by using the angular distribution of radial and ortho-radial forces $f^{\prime}_n$ and $f^{\prime}_t$, respectively. 
These angular distributions can be computed as 
\begin{eqnarray}
\langle f^{\prime}_n \rangle(\Theta^{\prime}) & = &\frac{1}{N_{c}(\Theta^{\prime})} \sum_{c \in \delta \Theta^{\prime}} f^{\prime}_n, \hspace{0.5cm} \mathrm{and} \\
\langle f^{\prime}_t \rangle(\Theta^{\prime}) & = &\frac{1}{N_{c}(\Theta^{\prime})} \sum_{c \in \delta \Theta^{\prime}} f^{\prime}_t.
\end{eqnarray}

\begin{figure}
    \centering
    \subfigure {\label{fig:afn}(a)
    \includegraphics[width=0.45\columnwidth]{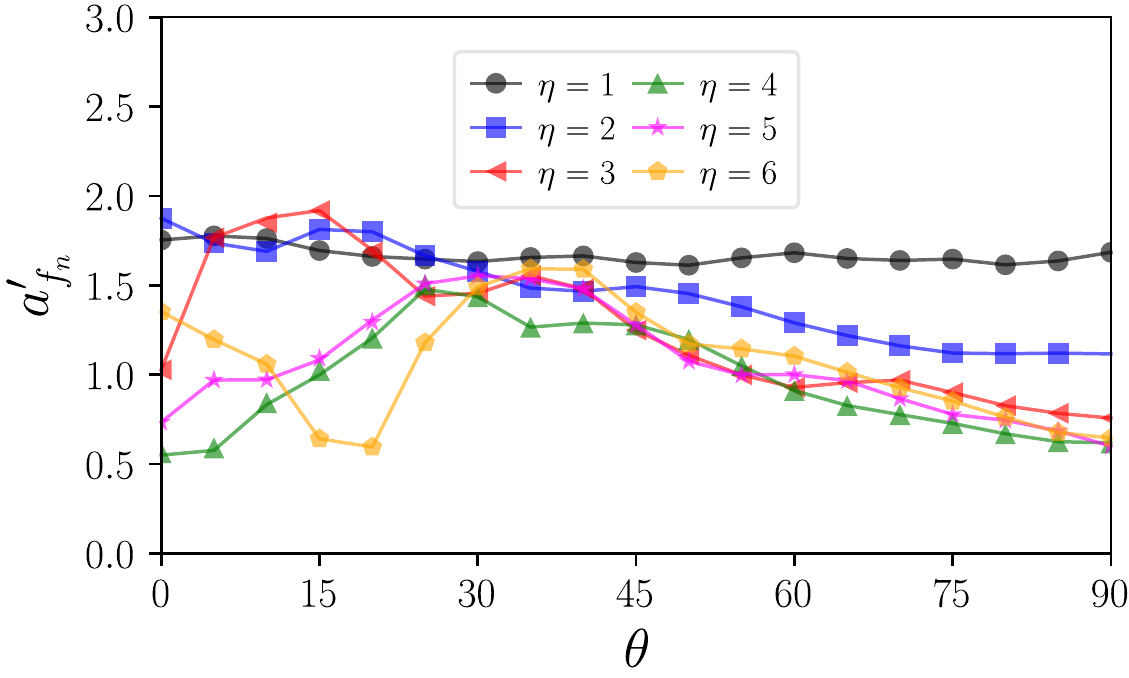}}
    \subfigure {\label{fig:t_afn}(b)
    \includegraphics[width=0.45\columnwidth]{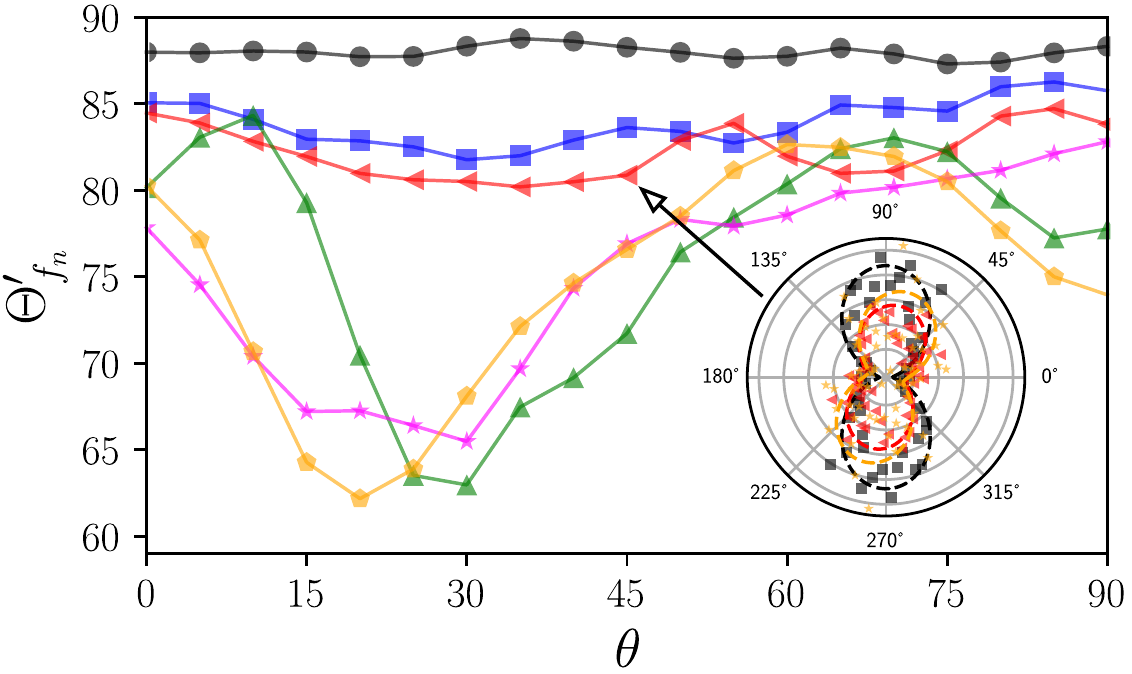}}
    \subfigure {\label{fig:aft}(c)
    \includegraphics[width=0.45\columnwidth]{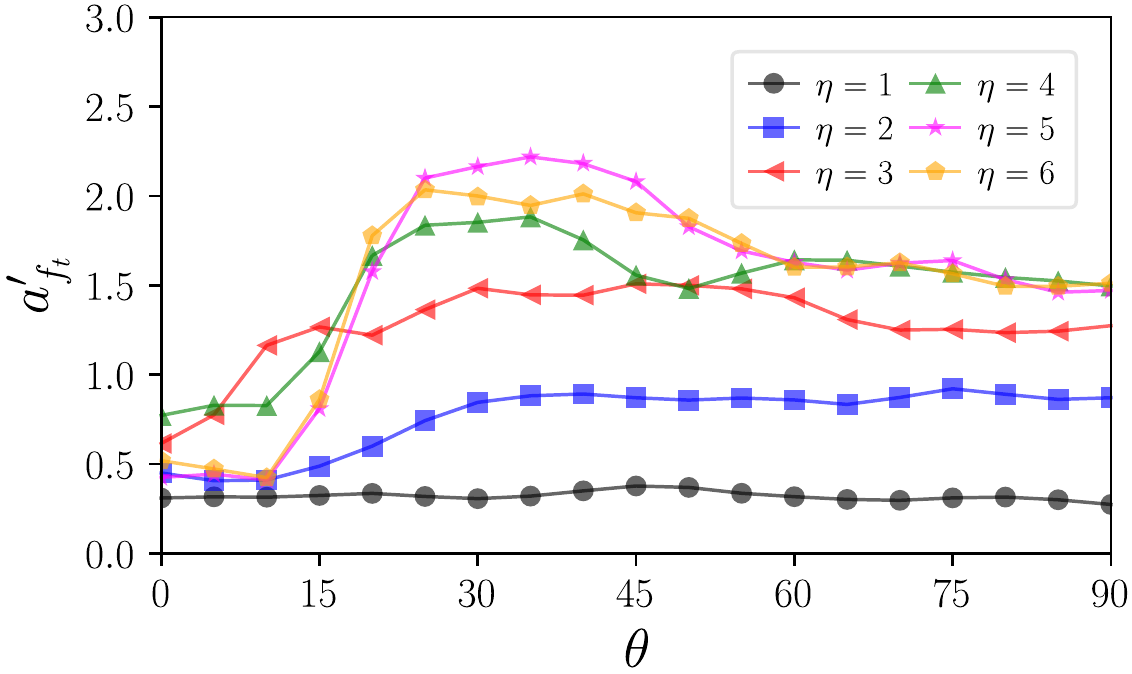}}
    \subfigure {\label{fig:t_aft}(d)
    \includegraphics[width=0.45\columnwidth]{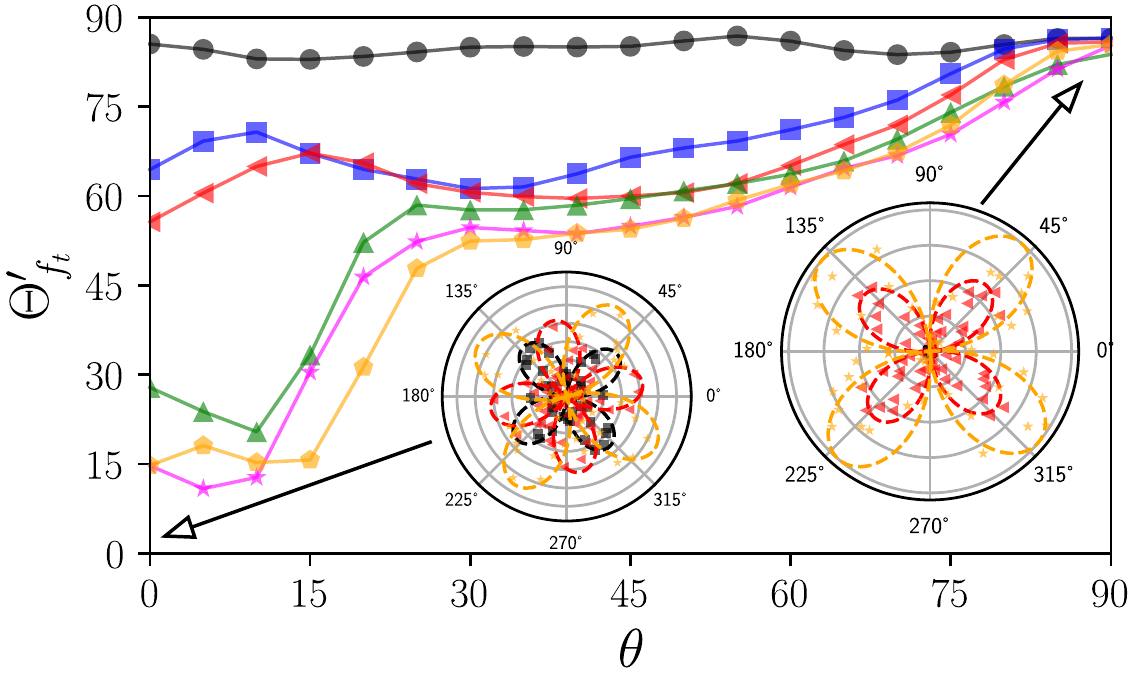}}
    \caption{Evolution of normal (top) and tangential (bottom) force anisotropies (left) and the preferential orientation of their respective angular distribution (right) for different values of inherent anisotropy $\eta$ and loading orientation $\theta$.}
    \label{fig:aniso_forces}
\end{figure}

In the insets of Fig. \ref{fig:t_afn}, we present the angular distribution $\langle f^{\prime}_n \rangle (\Theta^{\prime})$ for loading orientation $\theta=45^{\circ}$ which shows the misalignment of the largest forces with respect to the vertical (i.e., the loading orientation). 
In the insets of Fig. \ref{fig:t_aft}, we present the distributions of ortho-radial forces for $\theta = 0^{\circ}$ and $\theta = 90^{\circ}$, highlighting how widely these distributions vary as the assembly rotates. 
Regardless, these angular distributions remain periodic and smooth enough to fit Fourier series for their characterization. 
We can thus describe the angular variation of forces as 
\begin{eqnarray}\label{eq:f_n}
\langle f^{\prime}_n \rangle (\Theta^{\prime}) & = & \langle f^{\prime}_n \rangle \left\{1 + a^{\prime}_{f_n} \cos2\left(\Theta^{\prime}-\Theta^{\prime}_{f_n}\right) \right\}, \hspace{0.5cm} \mathrm{and} \\ 
\langle f^{\prime}_t \rangle (\Theta^{\prime}) & = & \langle f^{\prime}_n \rangle \left\{-a^{\prime}_{f_t} \sin2\left(\Theta^{\prime}-\Theta^{\prime}_{f_t}\right) \right\},
\end{eqnarray}
with  $a^{\prime}_{f_n}$ and $a^{\prime}_{f_t}$ being the level of anisotropy for each distribution, and $\Theta^{\prime}_{f_n}$ and $\Theta^{\prime}_{f_t}$ the respective preferential orientations.
For convenience, we build \emph{force tensors} that allow us to easily compute the anisotropies and main orientation of each distribution as 
\begin{eqnarray}
H^{f^{\prime}_n}_{ij} & = & \int_0^{\pi} \langle f^{\prime}_n \rangle(\Theta^{\prime}) n^{\prime}_i(\Theta^{\prime}) n_j(\Theta^{\prime}) d\Theta^{\prime}, \hspace{0.5cm} \mathrm{and}\\
H^{f^{\prime}_t}_{ij} & = & \int_0^{\pi} \langle f^{\prime}_t \rangle(\Theta^{\prime}) n_i(\Theta^{\prime}) t_j(\Theta^{\prime}) d\Theta^{\prime}.
\end{eqnarray}
This lets us compute the levels of force anisotropy as $a^{\prime}_{f_n} = 2\left(H^{f^{\prime}_n}_1 - H^{f^{\prime}_n}_2\right)/\left(H^{f^{\prime}_n}_1 + H^{f^{\prime}_n}_2\right)$ for the radial forces and $a^{\prime}_{f_t} = 2\left(H^{f^{\prime}_t}_1 - H^{f^{\prime}_t}_2\right)/\left(H^{f^{\prime}_n}_1 + H^{f^{\prime}_n}_2\right)$ for the ortho-radial forces, where $H^{\alpha}_1$ and $H^{\alpha}_2$ are the eigenvalues of each one of the tensors. 
Note that $H^{\alpha}_1 > H^{\alpha}_2$, and $\alpha$ stands for either the radial or ortho-radial components of the forces. 
It is worth mentioning that $tr\left(H^{f^{\prime}_n}\right) = \langle f^{\prime}_n \rangle$, i.e., the average radial force, and $tr\left(H^{f^{\prime}_t}\right) = 0$ by equilibrium of force moments over the cells. 

Figures \ref{fig:afn} and \ref{fig:t_afn} display the radial force anisotropies and their preferential orientations for the different levels of inherent anisotropy $\eta$ and loading orientation $\theta$.
In the case of $\eta = 1$, we can observe that $a^{\prime}_{f_n} \simeq 1.8$ independently of the loading orientation. 
Then, $a^{\prime}_{f_n}$ progressively increases with $\eta$ in the range $\theta \simeq [0^{\circ}, 30^{\circ}]$. 
After the loading orientation $\theta \simeq 30$, the radial force anisotropy presents a decreasing trend relatively similar among the different values of $\eta >1$. 
The preferred orientations of these forces show larger variations with $\theta$; a behavior that seems amplified with $\eta$. 
Also, note that for $\Theta^{\prime}_{f_n}$ a minimum value systematically appears around $\simeq 20^{\circ}$. 

For ortho-radial forces, Figs. \ref{fig:aft} and \ref{fig:t_aft} present the evolution of the level of anisotropy and preferential orientations as a function of $\eta$ and $\theta$. 
In this case, we observe a continuous increase of $a^{\prime}_{f_t}$ as a function of $\eta$ from loading orientations $\theta = 0^{\circ}$ to $\theta \simeq 45^{\circ}$.
Beyond that loading orientation, the ortho-radial force anisotropy reaches a plateau and barely varies with $\theta$. 
For the orientations $\Theta^{\prime}_{f_t}$, we can see a variation that becomes more important as the level of inherent anisotropy increases. 
Although we might have expected a joint evolution of the preferential radial and ortho-radial forces' orientations, these figures show that a non-evident trade-off of force anisotropies occurs for highly anisotropic materials. 
This fact is, of course, emphasized by the point loading configuration which signs the force transmission at bonds. 

\section{Scaling up the strength from the microstructure}\label{Sec:Model}
\subsection{Microstructural contributions to the stress tensor}
The previous microstructural parameters - concerning bonds, branches, and forces - must act together to produce the macroscopic failure strength we initially measured. 
This mapping between the micro and macro scales is especially challenging because of the varying shapes and sizes of the cells and the fact that the different microstructural tensors are strongly misaligned and evolving with $\theta$. 

In order to reconcile the micro and macro scales, let us consider the granular stress tensor as \citep{Andreotti2013, Nicot2013} 
\begin{equation}
\sigma_{ij} = \frac{1}{V} \sum_{\forall c} f^c_i \ell^c_j, 
\end{equation}
where $V$ is the volume of the sample, and the sum includes the dyadic product of the force $\bm{f}$ and branch $\bm{\ell}$ vectors for all interactions $c$.
Supposing that the distribution of forces and branches is uncorrelated (which is verified in our simulations), we can rewrite the stress tensor in terms of angular distributions on the frame $\{\bm{n^{\prime}},\bm{t^{\prime}}\}$ as \citep{Rothenburg1989}
\begin{equation}
\sigma_{ij} = n_c \int_0^{\pi} \left\{\langle f^{\prime}_n \rangle (\Theta^{\prime}) n^{\prime}_i(\Theta^{\prime}) - \langle f^{\prime}_t \rangle(\Theta^{\prime}) t^{\prime}_i(\Theta^{\prime}) \right\} \langle \ell \rangle(\Theta^{\prime}) n^{\prime}_j(\Theta^{\prime}) P_{c}(\Theta^{\prime}) d\Theta^{\prime},
\end{equation}
with $n_c$ being the bond number density defined as $N_c/V = Z/(2\langle V_{cl}\rangle)$, where $\langle V_{cl} \rangle$ is the average volume per cell. 
Note that we can also write $\langle V_{cl} \rangle = (\pi/4) \langle d_{cl} \rangle^2$, with $\langle d_{cl} \rangle$ being the equivalent average diameter of the cells. 
When we replace Eqs. (\ref{eq:P_c}), (\ref{eq:ell_n}), and (\ref{eq:f_n}) in the previous expression, focus only on the vertical component of the tensor (i.e., $\sigma_{yy}$), and integrate over the interval $[0,\pi]$, we find a microstructural definition of the vertical stress at the onset of failure as 
\begin{equation}\label{eq:micro_model}
\sigma^{th}_{yy} = \frac{Z \langle f^{\prime}_n \rangle \langle \ell \rangle}{\pi \langle d_{cl}\rangle^2} \left\{1 -\frac{1}{2}\sum a^{\prime}_k \cos2 \left( \Theta^{\prime}_k \right) + \frac{1}{2} \sum a^{\prime}_l a^{\prime}_m \cos2 \left(\Theta^{\prime}_m - \Theta^{\prime}_l\right) + \mathcal{O}\right\}.
\end{equation}

The term in brackets shows the contributions of the different anisotropies to the strength. 
In that term, the first sum runs in the set $a^{\prime}_k \in \{a^{\prime}_c,a^{\prime}_{\ell}, a^{\prime}_{f_n}, a^{\prime}_{f_t}\}$, and in the respective values for $\Theta^{\prime}_k$. 
The second sum is a product of anisotropies in which the combinations of indices $l$ and $m$ belong, respectively, to the set $\{a^{\prime}_{f_{n}} a^{\prime}_{\ell},a^{\prime}_{f_{n}} a^{\prime}_c, a^{\prime}_{\ell} a^{\prime}_c\}$, with the respective angles for $\Theta^{\prime}_m$ and $\Theta^{\prime}_l$. 
The higher-order term $\mathcal{O}$ involves triple products of anisotropies and is purposely neglected for the sake of simplicity.
Also note that we added the superscript `$th$' to emphasize that this value of strength results from the theoretical decomposition of the stress tensor. 
For simplicity, the term related to the anisotropies is henceforth written as $\mathcal{A}$. 

Equation (\ref{eq:micro_model}) illuminates the fact that non-trivial microstructural compensations occur between 1) the different anisotropy levels, 2) the preferred orientations of angular distributions, and 3) geometrical and mechanical features. 
In addition, the choice of the branch frame $\{\bm{n^\prime},\bm{t^\prime}\}$ instead of the bond frame $\{\bm{n},\bm{t}\}$ is deliberate because it allowed us to reduce the number of anisotropies and the number of terms involved in $\mathcal{A}$ \citep{Azema2010}. 

\begin{figure}
    \centering
    \subfigure {\label{fig:T1}
    \includegraphics[width=0.23\columnwidth]{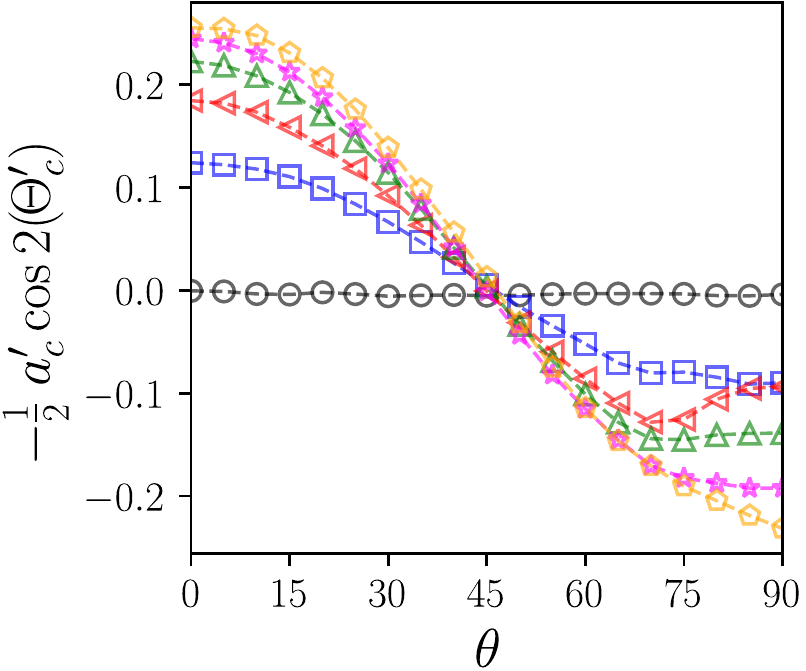}}
    \subfigure {\label{fig:T2}
    \includegraphics[width=0.23\columnwidth]{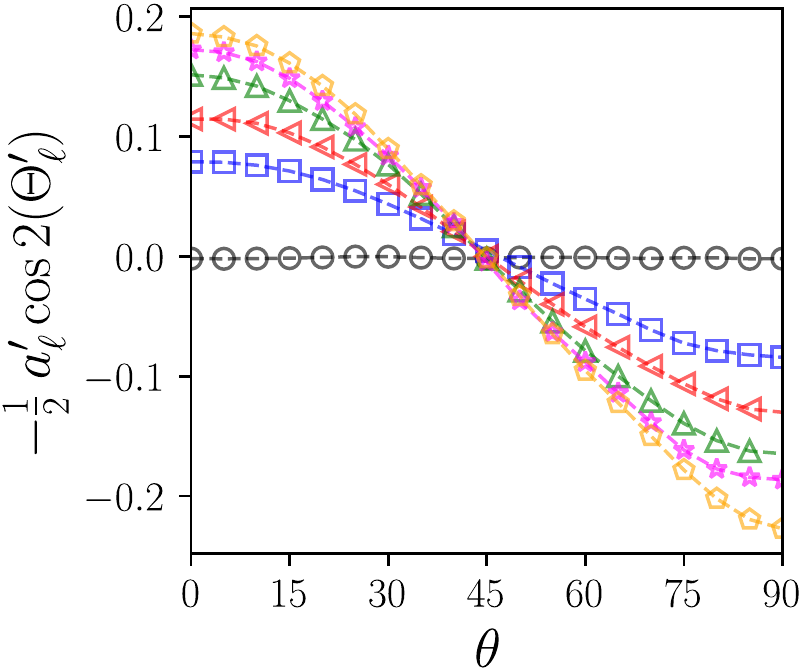}}
    \subfigure {\label{fig:T3}
    \includegraphics[width=0.23\columnwidth]{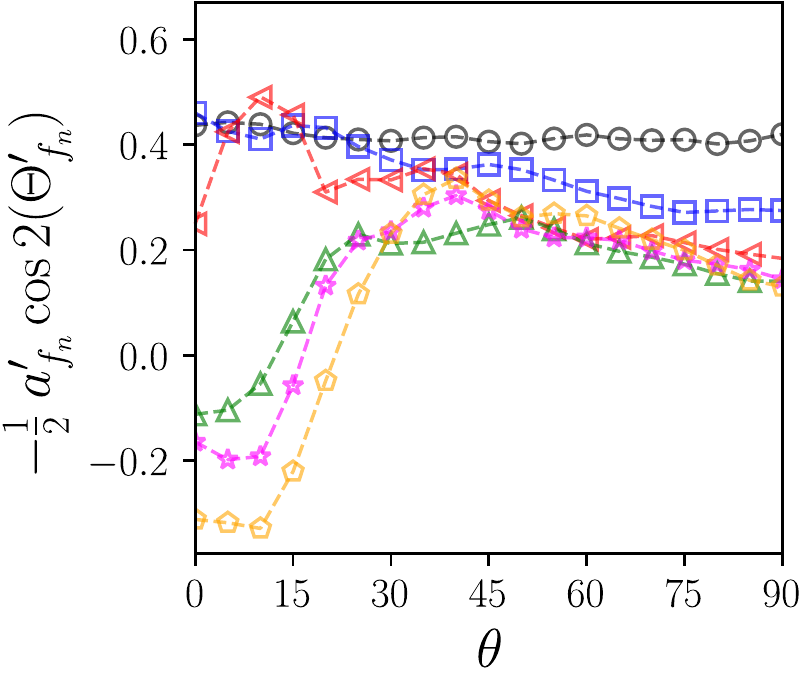}}
    \subfigure {\label{fig:T4}
    \includegraphics[width=0.23\columnwidth]{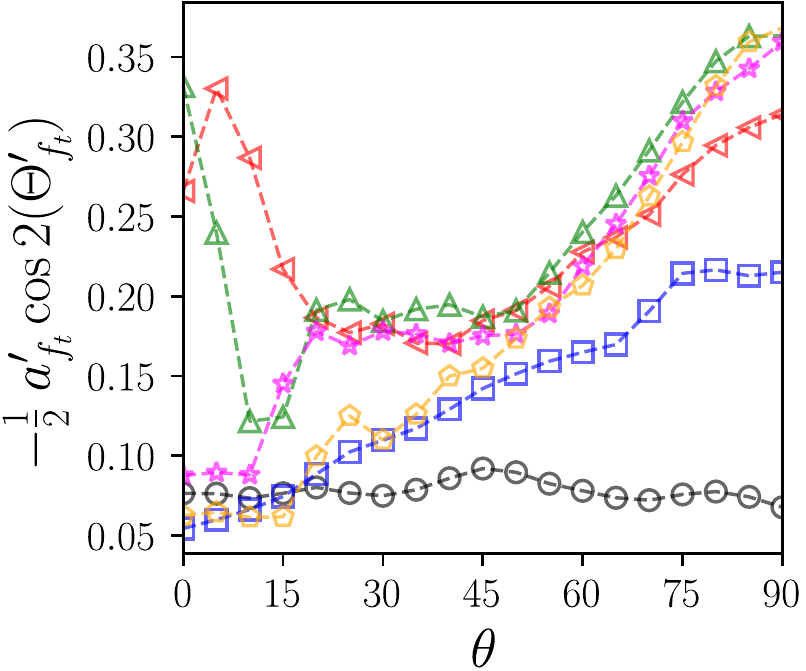}}
    \subfigure {\label{fig:T5}
    \includegraphics[width=0.23\columnwidth]{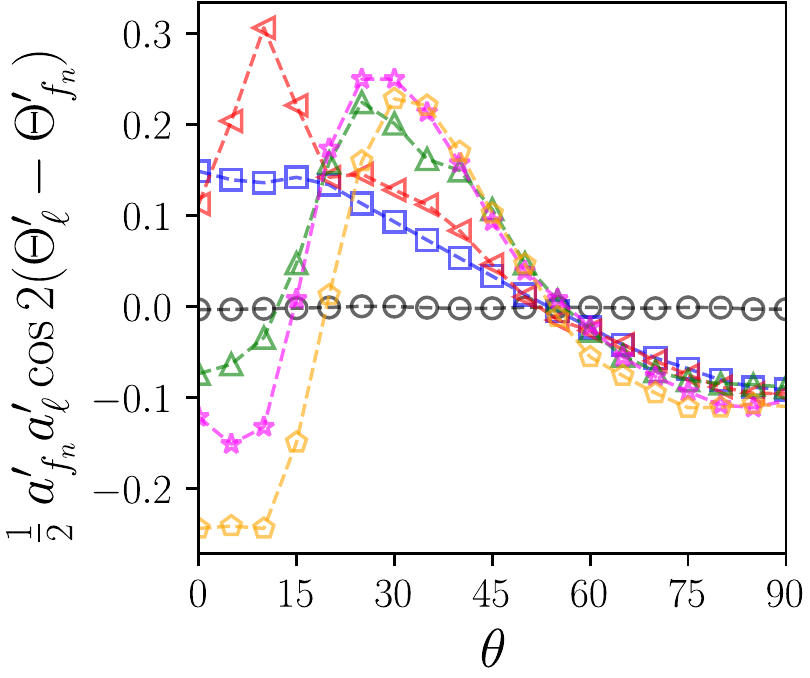}}
    \subfigure {\label{fig:T6}
    \includegraphics[width=0.23\columnwidth]{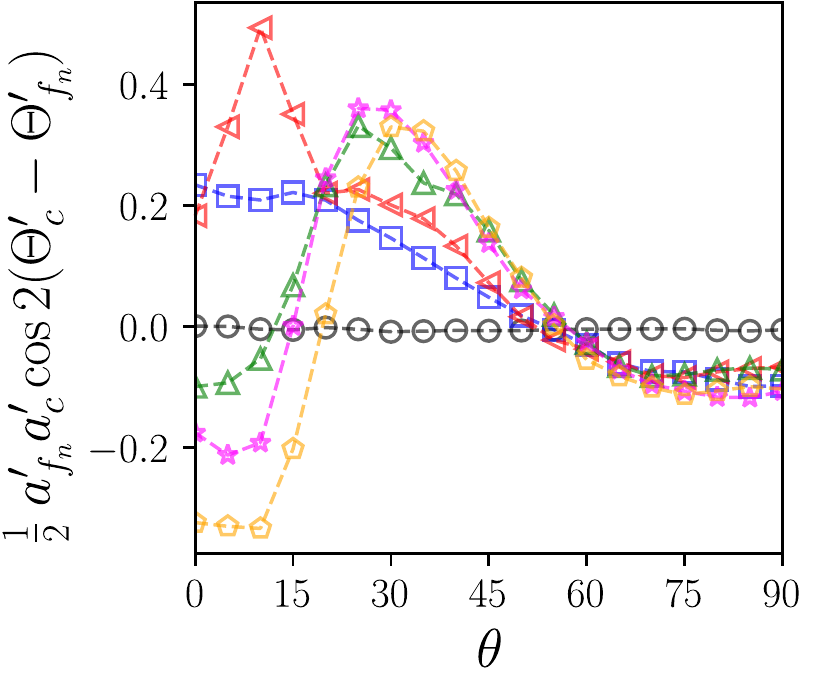}}
    \subfigure {\label{fig:T7}
    \includegraphics[width=0.23\columnwidth]{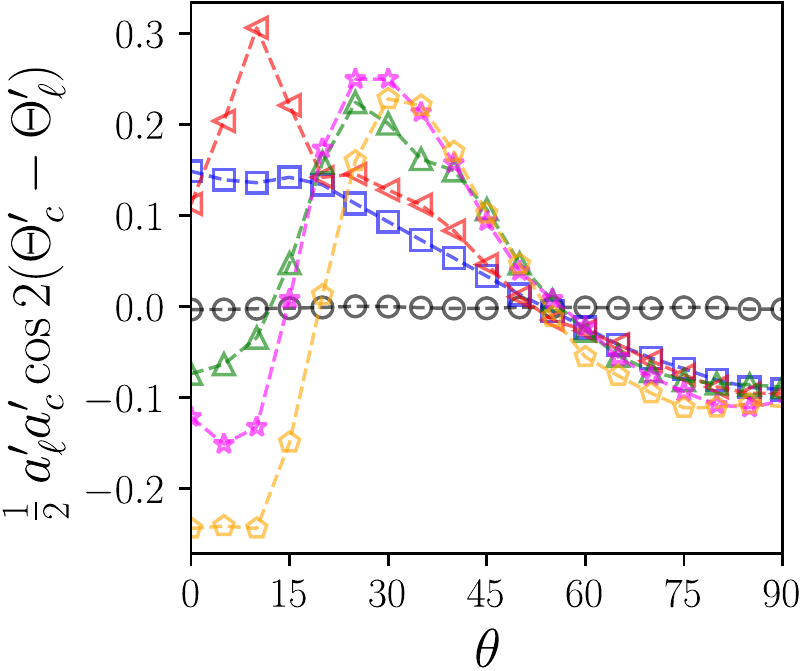}} \\
    \subfigure {\label{fig:A1}
    \includegraphics[width=0.5\columnwidth]{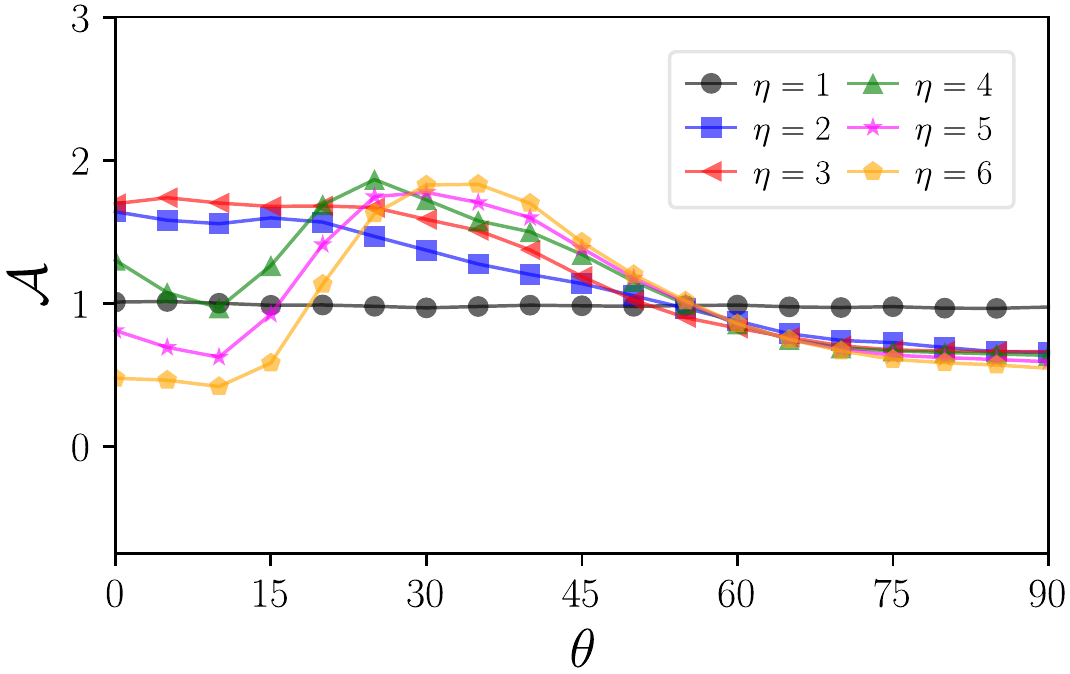}}
    \caption{Evolution of the terms in Eq. (\ref{eq:micro_model}) related to single anisotropies (first row) and double anisotropies (second row). 
    We also present the sum of these different parameters in the term $\mathcal{A}$ (bottom).}
    \label{fig:contributions}
\end{figure}

In Fig. \ref{fig:contributions}, we summarize the evolution of the different parameters involved in $\mathcal{A}$ for single and double anisotropies. 
On the one hand, we see that the geometrical anisotropies related to the branch orientation $a^{\prime}_c$ and branch lengths $a^{\prime}_{\ell}$ smoothly decrease as a function of the loading orientation $\theta$. 
On the other, the anisotropies related to the force transmission $a^{\prime}_{f_n}$ and $a^{\prime}_{f_t}$ have a highly non-linear evolution with $\theta$. 
For angles between $\theta = 0^{\circ}$ and $\theta = 30^{\circ}$, the radial force anisotropy increases but then finds a relatively steady value for larger loading orientations. 
For angles greater than $\theta \simeq 45^{\circ}$, the ortho-radial anisotropy increases strongly as a function of $\theta$. 
For the terms involving the product of anisotropies, the trends are all quite similar and not negligible in contribution. 

At the bottom of Fig. \ref{fig:contributions}, we see how all of these anisotropies add up.
Given the strong variation of all the anisotropies and preferential orientations, it is notable that the term $\mathcal{A}$ ends up fluctuating around the case $\eta=1$. 
This is clearly a mechanism involving direct compensations between geometrical microstructural characteristics and the force transmission at bonds. 
This phenomenon - in which the term $\mathcal{A}$ lies close to one - shows that the strong variation of the macroscopic failure strength must lie on the parameters $Z \langle f^{\prime}_n \rangle \langle \ell \rangle$ of the microstructural decomposition of stresses. 
Note that a version of Eq. (\ref{eq:micro_model}) that neglects the term $\mathcal{A}$ has been used many times before for conglomerates or granular assemblies in which particles are of similar size and shape \citep{Rumpf1970, Groger2003, Richefeu2006}. 
Nonetheless, as we just observed, $\mathcal{A}$ cannot be neglected for ellongated bodies. 

\begin{figure}
    \centering
    \subfigure {\label{fig:ln}(a)
    \includegraphics[width=0.45\columnwidth]{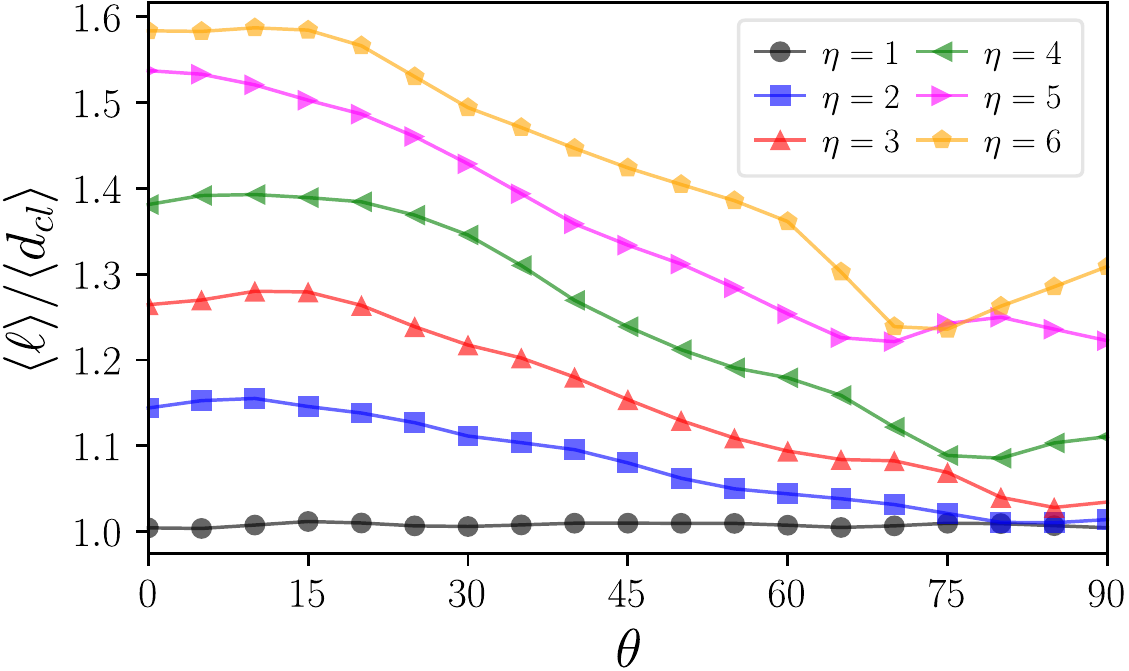}}
    \subfigure {\label{fig:fn}(b)
    \includegraphics[width=0.45\columnwidth]{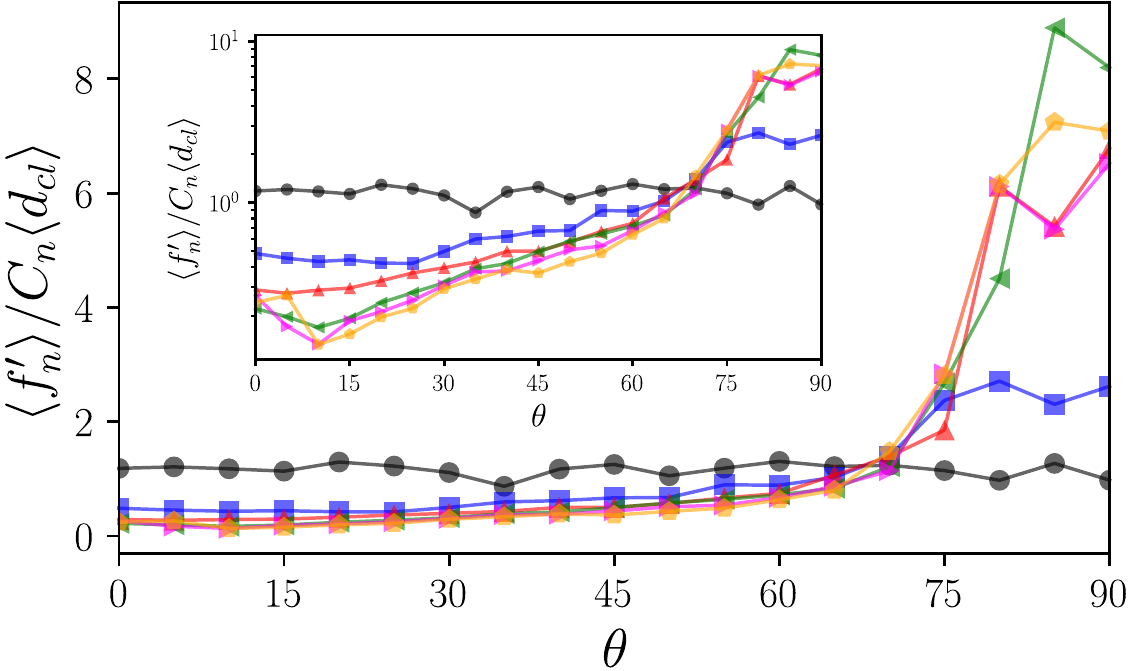}}
    
    \caption{(a) Evolution of the average branch length as a function of $\eta$ and $\theta$.
    (b) Evolution of the radial average force as a function of $\eta$ and $\theta$. In the inset: the same data in lin-log scale.}
    \label{fig:fn_ln}
\end{figure}

In Fig. \ref{fig:ln}, we see the evolution of the average branch length at the onset of failure as a function of the inherent anisotropy and the loading orientation, which shows a gradual drop as the loading becomes perpendicular to the layering. 
Such a variation is accentuated as $\eta$ grows. 
In Fig. \ref{fig:fn}, we present the evolution of the average radial force $\langle f^{\prime}_n \rangle$, which is normalized by the internal cohesion and the average cell equivalent diameter. 
This curve varies widely and, indeed, carries most of the shape of the macroscopic failure strength. 

These observations allow us to conclude that the microstructural mechanisms producing the increase of failure strength with $\theta$ are related to the rise of radial forces, the drop of average branch length and coordination number $Z$, and the complex compensations occurring within the term $\mathcal{A}$. 

\begin{figure}
    \centering
    \includegraphics[width=0.5\columnwidth]{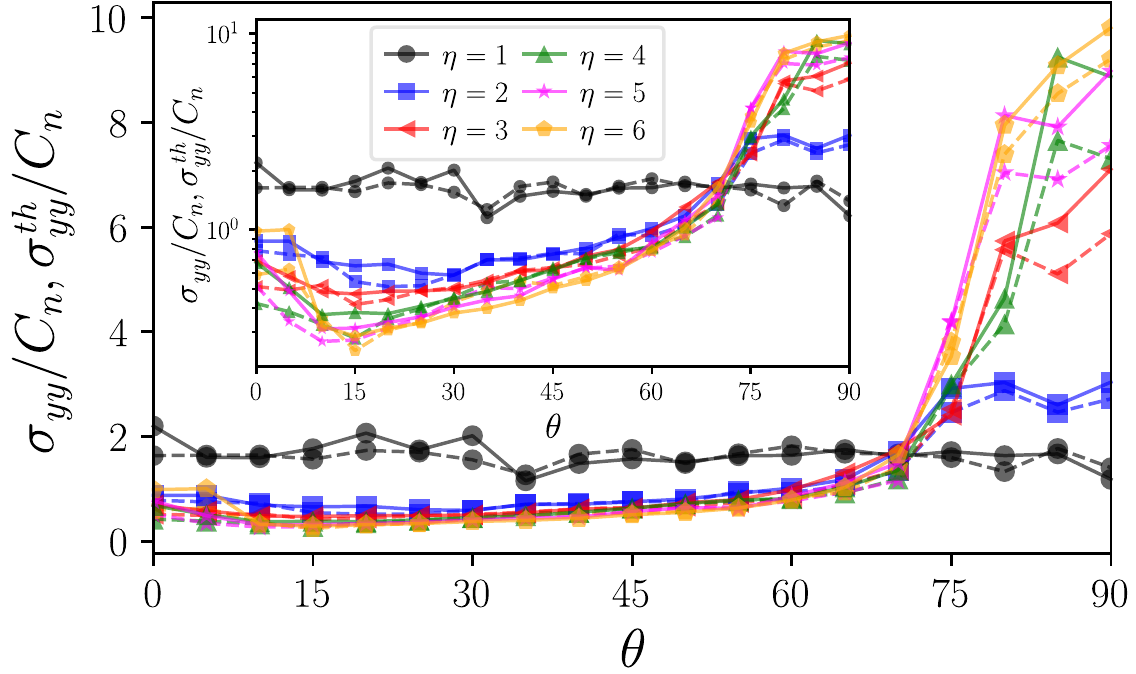}
    \caption{ Failure strength measure through the wall forces just as in Fig. \ref{fig:crushing_strength} (solid lines), and the same strength found using the microstructural decomposition of the stress tensor using Eq. (\ref{eq:sigma_yy}) (dashed lines).}
    \label{fig:mapping}
\end{figure}

Finally, in Fig. \ref{fig:mapping}, we plot $\sigma^{th}_{yy}$ nicely reproducing the macroscopic vertical failure stress measured in Sec. \ref{Sec:Macro}.
The small differences between the measure and the decomposition are linked to the higher-order terms that were neglected in Eq. (\ref{eq:micro_model}). 
Thus, based on a fine description of the phenomena at bonds and mineral organization in space, our micromechanical description proves capable of describing and scaling up the macroscopic behavior we observe in laboratory. 
%

\section{Summary}
We developed a series of numerical tests to study the failure strength of brittle materials reminiscent of schists, slates, shales, etc, whose components have a preferential orientation, i.e., an inherent anisotropy. 
Using a bi-dimensional discrete-element method, we built samples in which we could control the degree of inherent anisotropy by using a modified Voronoi tessellation. 
This approach allowed us to generate a set of subdivisions (the tessellation) of adjacent irregular polygons that we called cells. 
The common edges between cells interacted via cohesive bonds, enabling us to control both the failure strength and the cumulated surface energy necessary to produce fissuring. 
We then measured the macroscopic failure strength by applying a diametrical point load onto circular samples up to breakage. 
The failure strength turned out to be strongly affected by the layering orientation with respect to the loading direction $\theta$. 
As observed in many experimental tests, the failure strength in our numerical tests evolved in a parabolic `\emph{U}' shape, with a minimum value around a loading orientation of $\theta \simeq 25^{\circ}$.

We also analyzed the variability of the failure strength using the Weibull survival probability, concluding that mixing several anisotropic configurations and loading orientations may lead to misleading conclusions upon the average failure strength and data scatter. 
This means that experimental testing must consider the microstructure of samples to avoid a misinterpretation of the strength of anisotropic brittle materials. 

Finally, we performed a thorough characterization of geometrical properties of the cells' assemblies and force transmission mechanisms by means of the fabric, branch, and force tensors, as well as an approximation of their angular distributions using Fourier series. 
Exploiting the definition of the granular stress tensor in terms of angular contributions, we were able to find the microstructural elements that explain the variability of strength at the macroscopic scale. 
This was not a straightforward task. 
The strong geometrical and force anisotropies we found - as well as the misalignment of  the different tensors - prompted us to undertake a full description of the contributions of anisotropies involving high-order terms seldom seen when analyzing rocks or granular media. 
Instead of simplifying particles' shape and size variability, we modelled the complexity of these materials in order to identify the microstructural elements responsible for the macroscopic phenomena. 
We found that geometrical and mechanical anisotropies present complex compensations, which means they are not the main source of the failure strength variations. 
Rather, it is the cell coordination, the average branch length, and the average radial forces that present the larger fluctuations - making them the key microstructural elements at the origin of the macroscopic failure strength. 

Real materials are tremendously complex. 
Through this work, we sought to explore this complexity with the most detailed parameters we could gather linked to the granular stress tensor. 
Note that the circular shape we used for the samples was simply a choice of configuration that allowed us to compare our results to those obtained in rock  testing. 
Our approach, however, is general and can be extended to any sample shape, assemblies of many crushable bodies, and diverse bonding behavior other than pure cohesive. 
Many questions remain unresolved concerning the compaction or shear properties (rheology) of assemblies composed of several crushable inherently anisotropic bodies, which would benefit from future research. 

\section*{Acknowledgments}
This research work benefited from the financial support of the Natural Sciences and Engineering Research Council of Canada (NSERC) [Ref. RGPIN-2019-06118], the Fonds de recherche du Qu\'{e}bec - Nature et technologies (FRQNT) through the Programme de recherche en partenariat sur le d\'{e}veloppement durable du secteur minier-II [Ref. 2020-MN-281267] and the industrial partners of the Research Institute on Mines and the Environment (RIME) UQAT-Polytechnique (irme.ca/en).
We also thank Franny McGill for her careful editing of the manuscript.

\bibliographystyle{cas-model2-names}

\bibliography{biblio}

\begin{thebibliography}{72}
\expandafter\ifx\csname natexlab\endcsname\relax\def\natexlab#1{#1}\fi
\providecommand{\url}[1]{\texttt{#1}}
\providecommand{\href}[2]{#2}
\providecommand{\path}[1]{#1}
\providecommand{\DOIprefix}{doi:}
\providecommand{\ArXivprefix}{arXiv:}
\providecommand{\URLprefix}{URL: }
\providecommand{\Pubmedprefix}{pmid:}
\providecommand{\doi}[1]{\href{http://dx.doi.org/#1}{\path{#1}}}
\providecommand{\Pubmed}[1]{\href{pmid:#1}{\path{#1}}}
\providecommand{\bibinfo}[2]{#2}
\ifx\xfnm\relax \def\xfnm[#1]{\unskip,\space#1}\fi
\bibitem[{Amadei(1996)}]{Amadei1996}
\bibinfo{author}{Amadei, B.}, \bibinfo{year}{1996}.
\newblock \bibinfo{title}{{Importance of anisotropy when estimating and
  measuring in situ stresses in rock}}.
\newblock \bibinfo{journal}{Int. J. Rock Mech. Min. Sci} \bibinfo{volume}{33},
  \bibinfo{pages}{293--325}.
\bibitem[{{Amir Reza Beyabanaki} et~al.(2009){Amir Reza Beyabanaki}, Jafari,
  {Omid Reza Biabanaki} and {Ronald Yeung}}]{Beyabanaki2009}
\bibinfo{author}{{Amir Reza Beyabanaki}, S.}, \bibinfo{author}{Jafari, A.},
  \bibinfo{author}{{Omid Reza Biabanaki}, S.}, \bibinfo{author}{{Ronald Yeung},
  M.}, \bibinfo{year}{2009}.
\newblock \bibinfo{title}{{Nodal-based three-dimensional discontinuous
  deformation analysis (3-D DDA)}}.
\newblock \bibinfo{journal}{Computers and Geotechnics} \bibinfo{volume}{36},
  \bibinfo{pages}{359--372}.
\bibitem[{Andreotti et~al.(2013)Andreotti, Forterre and
  Pouliquen}]{Andreotti2013}
\bibinfo{author}{Andreotti, B.}, \bibinfo{author}{Forterre, Y.},
  \bibinfo{author}{Pouliquen, O.}, \bibinfo{year}{2013}.
\newblock \bibinfo{title}{{Granular media: between fluid and solid}}.
\newblock \bibinfo{publisher}{Cambridge University press}.
\bibitem[{Az{\'{e}}ma and Radjai(2010)}]{Azema2010}
\bibinfo{author}{Az{\'{e}}ma, E.}, \bibinfo{author}{Radjai, F.},
  \bibinfo{year}{2010}.
\newblock \bibinfo{title}{{Stress-strain behavior and geometrical properties of
  packings of elongated particles}}.
\newblock \bibinfo{journal}{Phys. Rev. E} \bibinfo{volume}{81}.
\bibitem[{{Bagherzadeh Kh.} et~al.(2011){Bagherzadeh Kh.}, Mirghasemi and
  Mohammadi}]{BagherzadehKh2011}
\bibinfo{author}{{Bagherzadeh Kh.}, A.}, \bibinfo{author}{Mirghasemi, A.},
  \bibinfo{author}{Mohammadi, S.}, \bibinfo{year}{2011}.
\newblock \bibinfo{title}{{Numerical simulation of particle breakage of angular
  particles using combined DEM and FEM}}.
\newblock \bibinfo{journal}{Powder Technology} \bibinfo{volume}{205},
  \bibinfo{pages}{15--29}.
\bibitem[{Bard et~al.(2013)Bard, Anabal{\'{o}}n and Campa{\~{n}}a}]{Bard2013}
\bibinfo{author}{Bard, E.}, \bibinfo{author}{Anabal{\'{o}}n, M.E.},
  \bibinfo{author}{Campa{\~{n}}a, J.}, \bibinfo{year}{2013}.
\newblock \bibinfo{title}{{Waste Rock Behavior at High Pressures: Dimensioning
  High Waste Rock Dumps}}.
\newblock \bibinfo{journal}{Multiscale Geomechanics} ,
  \bibinfo{pages}{83--112}.
\bibitem[{Brzesowsky et~al.(2014)Brzesowsky, Hangx, Brantut and
  Spiers}]{Brzesowsky2014}
\bibinfo{author}{Brzesowsky, R.}, \bibinfo{author}{Hangx, S.},
  \bibinfo{author}{Brantut, N.}, \bibinfo{author}{Spiers, C.},
  \bibinfo{year}{2014}.
\newblock \bibinfo{title}{{Compaction creep of sands due to time-dependent
  grain failure: Effects of chemical environment, applied stress, and grain
  size}}.
\newblock \bibinfo{journal}{AGU: Journal of Geophysical Research, Solid Earth}
  \bibinfo{volume}{199}, \bibinfo{pages}{7521--7541}.
\bibitem[{Cantor et~al.(2017)Cantor, Az{\'{e}}ma, Sornay and
  Radjai}]{Cantor2016}
\bibinfo{author}{Cantor, D.}, \bibinfo{author}{Az{\'{e}}ma, E.},
  \bibinfo{author}{Sornay, P.}, \bibinfo{author}{Radjai, F.},
  \bibinfo{year}{2017}.
\newblock \bibinfo{title}{{Three-dimensional bonded-cell model for grain
  fragmentation}}.
\newblock \bibinfo{journal}{Computational Particle Mechanics}
  \bibinfo{volume}{4}, \bibinfo{pages}{441--450}.
\bibitem[{Cantor et~al.(2015)Cantor, Estrada and Az{\'{e}}ma}]{Cantor2015}
\bibinfo{author}{Cantor, D.}, \bibinfo{author}{Estrada, N.},
  \bibinfo{author}{Az{\'{e}}ma, E.}, \bibinfo{year}{2015}.
\newblock \bibinfo{title}{{Split-Cell Method for grain fragmentation}}.
\newblock \bibinfo{journal}{Computers and Geotechnics} \bibinfo{volume}{67},
  \bibinfo{pages}{150--156}.
\bibitem[{Cantor et~al.(2021)Cantor, Ovalle and Az{\'{e}}ma}]{Cantor2021}
\bibinfo{author}{Cantor, D.}, \bibinfo{author}{Ovalle, C.},
  \bibinfo{author}{Az{\'{e}}ma, E.}, \bibinfo{year}{2021}.
\newblock \bibinfo{title}{{Strength and energy consumption of inherently
  anisotropic rocks at failure}}, in: \bibinfo{booktitle}{EPJ Web of
  Conferences}.
\newblock \bibinfo{note}{Accepted}.
\bibitem[{Chen et~al.(1998)Chen, Pan and Amadei}]{Chen1998}
\bibinfo{author}{Chen, C.S.}, \bibinfo{author}{Pan, E.},
  \bibinfo{author}{Amadei, B.}, \bibinfo{year}{1998}.
\newblock \bibinfo{title}{{Determination of Deformability and Tensile Strength
  of Anisotropic Rock Using Brazilian Tests}}.
\newblock \bibinfo{journal}{Int. J. Rock Mech. Min. Sci} \bibinfo{volume}{35},
  \bibinfo{pages}{43--61}.
\bibitem[{Cho et~al.(2007)Cho, Martin and Sego}]{Cho2007}
\bibinfo{author}{Cho, N.}, \bibinfo{author}{Martin, C.}, \bibinfo{author}{Sego,
  D.}, \bibinfo{year}{2007}.
\newblock \bibinfo{title}{{A clumped particle model for rock}}.
\newblock \bibinfo{journal}{Int. J. Rock Mech. Min. Sci} \bibinfo{volume}{44},
  \bibinfo{pages}{997--1010}.
\bibitem[{Ciantia et~al.(2015)Ciantia, Arroyo, Calvetti and Gens}]{Ciantia2015}
\bibinfo{author}{Ciantia, M.}, \bibinfo{author}{Arroyo, M.},
  \bibinfo{author}{Calvetti, F.}, \bibinfo{author}{Gens, A.},
  \bibinfo{year}{2015}.
\newblock \bibinfo{title}{{An approach to enhance efficiency of DEM modelling
  of soils with crushable grains}}.
\newblock \bibinfo{journal}{Geotechnique} , \bibinfo{pages}{91--110}.
\bibitem[{Du et~al.(2006)Du, Emelianenko and Ju}]{Du2006}
\bibinfo{author}{Du, Q.}, \bibinfo{author}{Emelianenko, M.},
  \bibinfo{author}{Ju, L.}, \bibinfo{year}{2006}.
\newblock \bibinfo{title}{Convergence properties of the lloyd algorithm for
  computing the centrodial voronoi tessellations}.
\newblock \bibinfo{journal}{SIAM Journal on Numerical Analysis}
  \bibinfo{volume}{44}, \bibinfo{pages}{102--119}.
\bibitem[{Dubois et~al.(2018)Dubois, Acary and Jean}]{Dubois2018}
\bibinfo{author}{Dubois, F.}, \bibinfo{author}{Acary, V.},
  \bibinfo{author}{Jean, M.}, \bibinfo{year}{2018}.
\newblock \bibinfo{title}{{The Contact Dynamics method: A nonsmooth story}}.
\newblock \bibinfo{journal}{Comptes Rendus - Mecanique} \bibinfo{volume}{346},
  \bibinfo{pages}{247--262}.
\bibitem[{Dubois et~al.(2020)Dubois, Jean and et~al}]{LMGC90Web}
\bibinfo{author}{Dubois, F.}, \bibinfo{author}{Jean, M.},
  \bibinfo{author}{et~al}, \bibinfo{year}{2020}.
\newblock \bibinfo{title}{{LMGC90 wiki page}}.
\newblock
  \bibinfo{howpublished}{\url{https://git-xen.lmgc.univ-montp2.fr/lmgc90/lmgc90_user/wikis/home}}.
\newblock \bibinfo{note}{[Online; accessed 17-Jul-2020]}.
\bibitem[{Dubois et~al.(2011)Dubois, Jean, Renouf, Mozul, Martin and
  Bagn{\'{e}}ris}]{Dubois2011}
\bibinfo{author}{Dubois, F.}, \bibinfo{author}{Jean, M.},
  \bibinfo{author}{Renouf, M.}, \bibinfo{author}{Mozul, R.},
  \bibinfo{author}{Martin, A.}, \bibinfo{author}{Bagn{\'{e}}ris, M.},
  \bibinfo{year}{2011}.
\newblock \bibinfo{title}{{LMGC90}}, in: \bibinfo{booktitle}{10e colloque
  national en calcul des structures}, p. \bibinfo{pages}{8 p}.
\bibitem[{Einav(2007)}]{Einav2007}
\bibinfo{author}{Einav, I.}, \bibinfo{year}{2007}.
\newblock \bibinfo{title}{{Breakage mechanics-Part I: Theory}}.
\newblock \bibinfo{journal}{J. Mech. Phys. Solids} \bibinfo{volume}{55},
  \bibinfo{pages}{1274--1297}.
\bibitem[{Favier et~al.(2006)Favier, Lazarus and Leblond}]{Favier2006}
\bibinfo{author}{Favier, E.}, \bibinfo{author}{Lazarus, V.},
  \bibinfo{author}{Leblond, J.B.}, \bibinfo{year}{2006}.
\newblock \bibinfo{title}{Statistics of the deformation of the front of a
  tunnel-crack propagating in some inhomogeneous medium}.
\newblock \bibinfo{journal}{J. Mech. Phys. Solids} \bibinfo{volume}{54},
  \bibinfo{pages}{1449 -- 1478}.
\bibitem[{Gao and Stead(2014)}]{Gao2014}
\bibinfo{author}{Gao, F.Q.}, \bibinfo{author}{Stead, D.}, \bibinfo{year}{2014}.
\newblock \bibinfo{title}{{The application of a modified Voronoi logic to
  brittle fracture modelling at the laboratory and field scale}}.
\newblock \bibinfo{journal}{Int. J. Rock Mech. Min. Sci} \bibinfo{volume}{68},
  \bibinfo{pages}{1--14}.
\bibitem[{Garagon and {\c{C}}an(2010)}]{Garagon2010}
\bibinfo{author}{Garagon, M.}, \bibinfo{author}{{\c{C}}an, T.},
  \bibinfo{year}{2010}.
\newblock \bibinfo{title}{{Predicting the strength anisotropy in uniaxial
  compression of some laminated sandstones using multivariate regression
  analysis}}.
\newblock \bibinfo{journal}{Materials and Structures/Materiaux et
  Constructions} \bibinfo{volume}{43}, \bibinfo{pages}{509--517}.
\bibitem[{Gladkyy and Kuna(2017)}]{Gladkyy2017}
\bibinfo{author}{Gladkyy, A.}, \bibinfo{author}{Kuna, M.},
  \bibinfo{year}{2017}.
\newblock \bibinfo{title}{{DEM simulation of polyhedral particle cracking using
  a combined Mohr–Coulomb–Weibull failure criterion}}.
\newblock \bibinfo{journal}{Granular Matter} \bibinfo{volume}{19},
  \bibinfo{pages}{1--11}.
\bibitem[{Griffith(1921)}]{Griffith1921}
\bibinfo{author}{Griffith, A.A.}, \bibinfo{year}{1921}.
\newblock \bibinfo{title}{{The Phenomena of Rupture and Flow in Solids}}.
\newblock \bibinfo{journal}{Philosophical Transactions of the Royal Society A:
  Mathematical, Physical and Engineering Sciences} \bibinfo{volume}{221},
  \bibinfo{pages}{163--198}.
\bibitem[{Gröger et~al.(2003)Gröger, Tüzün and Heyes}]{Groger2003}
\bibinfo{author}{Gröger, T.}, \bibinfo{author}{Tüzün, U.},
  \bibinfo{author}{Heyes, D.M.}, \bibinfo{year}{2003}.
\newblock \bibinfo{title}{Modelling and measuring of cohesion in wet granular
  materials}.
\newblock \bibinfo{journal}{Powder Technology} \bibinfo{volume}{133},
  \bibinfo{pages}{203 -- 215}.
\bibitem[{{Guha Roy} and Singh(2016)}]{GuhaRoy2016}
\bibinfo{author}{{Guha Roy}, D.}, \bibinfo{author}{Singh, T.N.},
  \bibinfo{year}{2016}.
\newblock \bibinfo{title}{{Effect of Heat Treatment and Layer Orientation on
  the Tensile Strength of a Crystalline Rock Under Brazilian Test Condition}}.
\newblock \bibinfo{journal}{Rock Mechanics and Rock Engineering}
  \bibinfo{volume}{49}, \bibinfo{pages}{1663--1677}.
\bibitem[{Guo and Zhao(2014)}]{Guo2014}
\bibinfo{author}{Guo, N.}, \bibinfo{author}{Zhao, J.}, \bibinfo{year}{2014}.
\newblock \bibinfo{title}{{A coupled FEM/DEM approach for hierarchical
  multiscale modelling of granular media}}.
\newblock \bibinfo{journal}{International journal for numerical methods in
  engineering} \bibinfo{volume}{99}, \bibinfo{pages}{789--818}.
\bibitem[{Hiramatsu and Oka(1966)}]{Hiramatsu1966}
\bibinfo{author}{Hiramatsu, Y.}, \bibinfo{author}{Oka, Y.},
  \bibinfo{year}{1966}.
\newblock \bibinfo{title}{Determination of the tensile strength of rock by a
  compression test of an irregular test piece}.
\newblock \bibinfo{journal}{International Journal of Rock Mechanics and Mining
  Sciences \& Geomechanics Abstracts} \bibinfo{volume}{3}, \bibinfo{pages}{89
  -- 90}.
\bibitem[{Hoek(1964)}]{Hoek1964}
\bibinfo{author}{Hoek, E.}, \bibinfo{year}{1964}.
\newblock \bibinfo{title}{{Fracture of Anisotropic Rock}}.
\newblock \bibinfo{journal}{Journal of the South African Institute of Mining
  and Metallurgy} \bibinfo{volume}{64}, \bibinfo{pages}{501--518}.
\bibitem[{Huillca et~al.(2020)Huillca, Silva, Ovalle, Carrasco, Quezada and
  Villavicencio}]{Huillca2020}
\bibinfo{author}{Huillca, Y.}, \bibinfo{author}{Silva, M.},
  \bibinfo{author}{Ovalle, C.}, \bibinfo{author}{Carrasco, S.},
  \bibinfo{author}{Quezada, J.}, \bibinfo{author}{Villavicencio, G.},
  \bibinfo{year}{2020}.
\newblock \bibinfo{title}{{Modeling size effect on rock aggregates strength
  using a DEM bonded-cell model}}.
\newblock \bibinfo{journal}{Acta Geotechnica} .
\bibitem[{Hurley et~al.(2018)Hurley, Lind, Pagan, Akin and
  Herbold}]{Hurley2018}
\bibinfo{author}{Hurley, R.C.}, \bibinfo{author}{Lind, J.},
  \bibinfo{author}{Pagan, D.C.}, \bibinfo{author}{Akin, M.C.},
  \bibinfo{author}{Herbold, E.B.}, \bibinfo{year}{2018}.
\newblock \bibinfo{title}{{In situ grain fracture mechanics during uniaxial
  compaction of granular solids}}.
\newblock \bibinfo{journal}{J. Mech. Phys. Solids} \bibinfo{volume}{112},
  \bibinfo{pages}{273--290}.
\bibitem[{Iliev et~al.(2019)Iliev, Wittel and Herrmann}]{Iliev2019}
\bibinfo{author}{Iliev, P.S.}, \bibinfo{author}{Wittel, F.K.},
  \bibinfo{author}{Herrmann, H.J.}, \bibinfo{year}{2019}.
\newblock \bibinfo{title}{{Evolution of fragment size distributions from the
  crushing of granular materials}}.
\newblock \bibinfo{journal}{Phys. Rev. E} \bibinfo{volume}{99},
  \bibinfo{pages}{1--10}.
\bibitem[{Indraratna et~al.(2011)Indraratna, Salim and
  Rujikiatkamjorn}]{Indraratna2011}
\bibinfo{author}{Indraratna, B.}, \bibinfo{author}{Salim, W.},
  \bibinfo{author}{Rujikiatkamjorn, C.}, \bibinfo{year}{2011}.
\newblock \bibinfo{title}{Advanced rail geotechnology-ballasted track}.
\newblock \bibinfo{publisher}{CRS Press}.
\bibitem[{Jaeger(1967)}]{Jaeger1967}
\bibinfo{author}{Jaeger, J.}, \bibinfo{year}{1967}.
\newblock \bibinfo{title}{Failure of rocks under tensile conditions}.
\newblock \bibinfo{journal}{International Journal of Rock Mechanics and Mining
  Sciences \& Geomechanics Abstracts} \bibinfo{volume}{4}, \bibinfo{pages}{219
  -- 227}.
\bibitem[{Jaeger et~al.(2007)Jaeger, Cook and Zimmerman}]{Jaeger2007}
\bibinfo{author}{Jaeger, J.}, \bibinfo{author}{Cook, N.},
  \bibinfo{author}{Zimmerman, R.}, \bibinfo{year}{2007}.
\newblock \bibinfo{title}{Fundamentals of Rock Mechanics}.
\newblock \bibinfo{publisher}{Wiley}.
\bibitem[{Jean(1999)}]{Jean1999}
\bibinfo{author}{Jean, M.}, \bibinfo{year}{1999}.
\newblock \bibinfo{title}{{The non-smooth contact dynamics method}}.
\newblock \bibinfo{journal}{Computer Methods in Applied Mechanics and
  Engineering} \bibinfo{volume}{177}, \bibinfo{pages}{235--257}.
\bibitem[{Jones and Ashby(2019)}]{Jones2019}
\bibinfo{author}{Jones, D.R.}, \bibinfo{author}{Ashby, M.F.},
  \bibinfo{year}{2019}.
\newblock \bibinfo{title}{Engineering materials 1},
  \bibinfo{publisher}{Butterworth-Heinemann}.
\bibitem[{Karakul et~al.(2010)Karakul, Ulusay and Isik}]{Karakul2010}
\bibinfo{author}{Karakul, H.}, \bibinfo{author}{Ulusay, R.},
  \bibinfo{author}{Isik, N.S.}, \bibinfo{year}{2010}.
\newblock \bibinfo{title}{{Empirical models and numerical analysis for
  assessing strength anisotropy based on block punch index and uniaxial
  compression tests}}.
\newblock \bibinfo{journal}{Int. J. Rock Mech. Min. Sci} \bibinfo{volume}{47},
  \bibinfo{pages}{657--665}.
\bibitem[{Kazerani(2013)}]{Kazerani2013}
\bibinfo{author}{Kazerani, T.}, \bibinfo{year}{2013}.
\newblock \bibinfo{title}{{Effect of micromechanical parameters of
  microstructure on compressive and tensile failure process of rock}}.
\newblock \bibinfo{journal}{Int. J. Rock Mech. Min. Sci} \bibinfo{volume}{64},
  \bibinfo{pages}{44--55}.
\bibitem[{Khanlari et~al.(2015)Khanlari, Rafiei and Abdilor}]{Khanlari2015}
\bibinfo{author}{Khanlari, G.}, \bibinfo{author}{Rafiei, B.},
  \bibinfo{author}{Abdilor, Y.}, \bibinfo{year}{2015}.
\newblock \bibinfo{title}{{An Experimental Investigation of the Brazilian
  Tensile Strength and Failure Patterns of Laminated Sandstones}}.
\newblock \bibinfo{journal}{Rock Mechanics and Rock Engineering}
  \bibinfo{volume}{48}, \bibinfo{pages}{843--852}.
\bibitem[{Lan et~al.(2010)Lan, Martin and Hu}]{Lan2010}
\bibinfo{author}{Lan, H.}, \bibinfo{author}{Martin, C.D.}, \bibinfo{author}{Hu,
  B.}, \bibinfo{year}{2010}.
\newblock \bibinfo{title}{{Effect of heterogeneity of brittle rock on
  micromechanical extensile behavior during compression loading}}.
\newblock \bibinfo{journal}{Journal of Geophysical Research}
  \bibinfo{volume}{115}.
\bibitem[{Lim et~al.(2004)Lim, McDowell and Collop}]{Lim2004}
\bibinfo{author}{Lim, W.L.}, \bibinfo{author}{McDowell, G.R.},
  \bibinfo{author}{Collop, A.C.}, \bibinfo{year}{2004}.
\newblock \bibinfo{title}{{The application of Weibull statistics to the
  strength of railway ballast}}.
\newblock \bibinfo{journal}{Granular Matter} \bibinfo{volume}{6},
  \bibinfo{pages}{229--237}.
\bibitem[{Ma et~al.(2014)Ma, Zhou, Chang and Yuan}]{Ma2014}
\bibinfo{author}{Ma, G.}, \bibinfo{author}{Zhou, W.}, \bibinfo{author}{Chang,
  X.L.}, \bibinfo{author}{Yuan, W.}, \bibinfo{year}{2014}.
\newblock \bibinfo{title}{{Combined FEM/DEM modeling of triaxial compression
  tests for rockfills with polyhedral particles}}.
\newblock \bibinfo{journal}{International Journal of Geomechanics}
  \bibinfo{volume}{14}, \bibinfo{pages}{1--12}.
\bibitem[{Mahabadi et~al.(2010)Mahabadi, Cottrell and Grasselli}]{Mahabadi2010}
\bibinfo{author}{Mahabadi, O.K.}, \bibinfo{author}{Cottrell, B.E.},
  \bibinfo{author}{Grasselli, G.}, \bibinfo{year}{2010}.
\newblock \bibinfo{title}{{An example of realistic modelling of rock dynamics
  problems: FEM/DEM simulation of dynamic brazilian test on Barre Granite}}.
\newblock \bibinfo{journal}{Rock Mechanics and Rock Engineering}
  \bibinfo{volume}{43}, \bibinfo{pages}{707--716}.
\bibitem[{Main and Meredith(1991)}]{Main1991}
\bibinfo{author}{Main, I.G.}, \bibinfo{author}{Meredith, P.G.},
  \bibinfo{year}{1991}.
\newblock \bibinfo{title}{{Stress corrosion constitutive laws as a possible
  mechanism of intermediate‐term and short‐term seismic quiescence}}.
\newblock \bibinfo{journal}{Geophysical Journal International}
  \bibinfo{volume}{107}, \bibinfo{pages}{363--372}.
\bibitem[{Marinelli and Buscarnera(2019)}]{Marinelli2019}
\bibinfo{author}{Marinelli, F.}, \bibinfo{author}{Buscarnera, G.},
  \bibinfo{year}{2019}.
\newblock \bibinfo{title}{{Anisotropic breakage mechanics: From stored energy
  to yielding in transversely isotropic granular rocks}}.
\newblock \bibinfo{journal}{J. Mech. Phys. Solids} \bibinfo{volume}{129},
  \bibinfo{pages}{1--18}.
\bibitem[{Marsal(1973)}]{Marsal1973}
\bibinfo{author}{Marsal, R.}, \bibinfo{year}{1973}.
\newblock \bibinfo{title}{Mechanical properties of rockfill dams}.
\newblock \bibinfo{publisher}{ISTE Ltd and John Wiley \& Sons Inc.}
\bibitem[{McDowell and Bolton(1998)}]{McDowell1998}
\bibinfo{author}{McDowell, G.R.}, \bibinfo{author}{Bolton, M.D.},
  \bibinfo{year}{1998}.
\newblock \bibinfo{title}{{On the micromechanics of crushable aggregates}}.
\newblock \bibinfo{journal}{Geotechnique} \bibinfo{volume}{48},
  \bibinfo{pages}{667--679}.
\bibitem[{Nguyen et~al.(2015)Nguyen, Az{\'{e}}ma, Sornay and
  Radjai}]{Nguyen2015}
\bibinfo{author}{Nguyen, D.H.}, \bibinfo{author}{Az{\'{e}}ma, E.},
  \bibinfo{author}{Sornay, P.}, \bibinfo{author}{Radjai, F.},
  \bibinfo{year}{2015}.
\newblock \bibinfo{title}{{Bonded-cell model for particle fracture}}.
\newblock \bibinfo{journal}{Phys. Rev. E} \bibinfo{volume}{91},
  \bibinfo{pages}{022203}.
\bibitem[{Nicot et~al.(2013)Nicot, Hadda, Guessasma, Fortin and
  Millet}]{Nicot2013}
\bibinfo{author}{Nicot, F.}, \bibinfo{author}{Hadda, N.},
  \bibinfo{author}{Guessasma, M.}, \bibinfo{author}{Fortin, J.},
  \bibinfo{author}{Millet, O.}, \bibinfo{year}{2013}.
\newblock \bibinfo{title}{On the definition of the stress tensor in granular
  media}.
\newblock \bibinfo{journal}{Int. J. Solids Struct.} \bibinfo{volume}{50},
  \bibinfo{pages}{2508 -- 2517}.
\bibitem[{Oda(1982)}]{Oda1982}
\bibinfo{author}{Oda, M.}, \bibinfo{year}{1982}.
\newblock \bibinfo{title}{{Fabric tensor for discontinuous geological
  materials}}.
\newblock \bibinfo{journal}{Soils and Foundations} \bibinfo{volume}{22}.
\bibitem[{Orozco et~al.(2019)Orozco, Delenne, Sornay and Radjai}]{Orozco2019}
\bibinfo{author}{Orozco, L.F.}, \bibinfo{author}{Delenne, J.Y.},
  \bibinfo{author}{Sornay, P.}, \bibinfo{author}{Radjai, F.},
  \bibinfo{year}{2019}.
\newblock \bibinfo{title}{{Discrete-element model for dynamic fracture of a
  single particle}}.
\newblock \bibinfo{journal}{Int. J. Solids Struct.} \bibinfo{volume}{166},
  \bibinfo{pages}{47--56}.
\bibitem[{Ovalle et~al.(2014)Ovalle, Frossard, Dano, Hu, Maiolino and
  Hicher}]{Ovalle2014}
\bibinfo{author}{Ovalle, C.}, \bibinfo{author}{Frossard, E.},
  \bibinfo{author}{Dano, C.}, \bibinfo{author}{Hu, W.},
  \bibinfo{author}{Maiolino, S.}, \bibinfo{author}{Hicher, P.Y.},
  \bibinfo{year}{2014}.
\newblock \bibinfo{title}{{The effect of size on the strength of coarse rock
  aggregates and large rockfill samples through experimental data}}.
\newblock \bibinfo{journal}{Acta Mechanica} \bibinfo{volume}{225},
  \bibinfo{pages}{2199--2216}.
\bibitem[{Ovalle et~al.(2020)Ovalle, Linero, Dano, Bard, Hicher and
  Osses}]{Ovalle2020}
\bibinfo{author}{Ovalle, C.}, \bibinfo{author}{Linero, S.},
  \bibinfo{author}{Dano, C.}, \bibinfo{author}{Bard, E.},
  \bibinfo{author}{Hicher, P.Y.}, \bibinfo{author}{Osses, R.},
  \bibinfo{year}{2020}.
\newblock \bibinfo{title}{{Data Compilation from Large Drained Compression
  Triaxial Tests on Coarse Crushable Rockfill Materials}}.
\newblock \bibinfo{journal}{Journal of Geotechnical and Geoenvironmental
  Engineering} \bibinfo{volume}{146}, \bibinfo{pages}{06020013}.
\bibitem[{Ovalle et~al.(2016)Ovalle, Voivret, Dano and Hicher}]{Ovalle2016}
\bibinfo{author}{Ovalle, C.}, \bibinfo{author}{Voivret, C.},
  \bibinfo{author}{Dano, C.}, \bibinfo{author}{Hicher, P.Y.},
  \bibinfo{year}{2016}.
\newblock \bibinfo{title}{Population balance in confined comminution using a
  physically based probabilistic approach for polydisperse granular materials}.
\newblock \bibinfo{journal}{International Journal for Numerical and Analytical
  Methods in Geomechanics} \bibinfo{volume}{40}, \bibinfo{pages}{2383--2397}.
\bibitem[{Pindra et~al.(2010)Pindra, Lazarus and Leblond}]{Pindra2010}
\bibinfo{author}{Pindra, N.}, \bibinfo{author}{Lazarus, V.},
  \bibinfo{author}{Leblond, J.B.}, \bibinfo{year}{2010}.
\newblock \bibinfo{title}{Geometrical disorder of the fronts of a tunnel-crack
  propagating in shear in some heterogeneous medium}.
\newblock \bibinfo{journal}{J. Mech. Phys. Solids} \bibinfo{volume}{58},
  \bibinfo{pages}{281 -- 299}.
\bibitem[{Potyondy and Cundall(2004)}]{Potyondy2004}
\bibinfo{author}{Potyondy, D.O.}, \bibinfo{author}{Cundall, P.A.},
  \bibinfo{year}{2004}.
\newblock \bibinfo{title}{{A bonded-particle model for rock}}.
\newblock \bibinfo{journal}{Int. J. Rock Mech. Min. Sci} \bibinfo{volume}{41},
  \bibinfo{pages}{1329--1364}.
\bibitem[{Pouragha et~al.(2020)Pouragha, Eghbalian and Wan}]{Pouragha2020}
\bibinfo{author}{Pouragha, M.}, \bibinfo{author}{Eghbalian, M.},
  \bibinfo{author}{Wan, R.}, \bibinfo{year}{2020}.
\newblock \bibinfo{title}{{Micromechanical correlation between elasticity and
  strength characteristics of anisotropic rocks}}.
\newblock \bibinfo{journal}{Int. J. Rock Mech. Min. Sci} \bibinfo{volume}{125},
  \bibinfo{pages}{104154}.
\bibitem[{Radjai and Richefeu(2009)}]{Radjai2009}
\bibinfo{author}{Radjai, F.}, \bibinfo{author}{Richefeu, V.},
  \bibinfo{year}{2009}.
\newblock \bibinfo{title}{{Contact dynamics as a nonsmooth discrete element
  method}}.
\newblock \bibinfo{journal}{Mechanics of Materials} \bibinfo{volume}{41},
  \bibinfo{pages}{715--728}.
\bibitem[{Renouf et~al.(2004)Renouf, Dubois and Alart}]{Renouf2004}
\bibinfo{author}{Renouf, M.}, \bibinfo{author}{Dubois, F.},
  \bibinfo{author}{Alart, P.}, \bibinfo{year}{2004}.
\newblock \bibinfo{title}{{A parallel version of the non smooth contact
  dynamics algorithm applied to the simulation of granular media}}.
\newblock \bibinfo{journal}{Journal of Computational and Applied Mathematics}
  \bibinfo{volume}{168}, \bibinfo{pages}{375--382}.
\bibitem[{Richefeu et~al.(2006)Richefeu, {El Youssoufi} and
  Radjai}]{Richefeu2006}
\bibinfo{author}{Richefeu, V.}, \bibinfo{author}{{El Youssoufi}, M.S.},
  \bibinfo{author}{Radjai, F.}, \bibinfo{year}{2006}.
\newblock \bibinfo{title}{{Shear strength properties of wet granular
  materials}}.
\newblock \bibinfo{journal}{Phys. Rev. E} \bibinfo{volume}{73},
  \bibinfo{pages}{051304}.
\bibitem[{Rothenburg and Bathurst(1989)}]{Rothenburg1989}
\bibinfo{author}{Rothenburg, L.}, \bibinfo{author}{Bathurst, R.J.},
  \bibinfo{year}{1989}.
\newblock \bibinfo{title}{{Analytical study of induced anisotropy in idealized
  granular material}}.
\newblock \bibinfo{journal}{G{\'{e}}otechnique} \bibinfo{volume}{39},
  \bibinfo{pages}{601--614}.
\bibitem[{Rumpf(1970)}]{Rumpf1970}
\bibinfo{author}{Rumpf, H.C.H.}, \bibinfo{year}{1970}.
\newblock \bibinfo{title}{Zur theorie der zugfestigkeit von agglomeraten bei
  kraftübertragung an kontaktpunkten}.
\newblock \bibinfo{journal}{Chemie Ingenieur Technik} \bibinfo{volume}{42},
  \bibinfo{pages}{538--540}.
\bibitem[{Saeidi et~al.(2014)Saeidi, Rasouli, Vaneghi, Gholami and
  Torabi}]{Saeidi2014}
\bibinfo{author}{Saeidi, O.}, \bibinfo{author}{Rasouli, V.},
  \bibinfo{author}{Vaneghi, R.G.}, \bibinfo{author}{Gholami, R.},
  \bibinfo{author}{Torabi, S.R.}, \bibinfo{year}{2014}.
\newblock \bibinfo{title}{{A modified failure criterion for transversely
  isotropic rocks}}.
\newblock \bibinfo{journal}{Geoscience Frontiers} \bibinfo{volume}{5},
  \bibinfo{pages}{215--225}.
\bibitem[{Sammis et~al.(1987)Sammis, King and Biegel}]{Sammis1987}
\bibinfo{author}{Sammis, C.}, \bibinfo{author}{King, G.},
  \bibinfo{author}{Biegel, R.}, \bibinfo{year}{1987}.
\newblock \bibinfo{title}{{The kinematics of gouge deformation}}.
\newblock \bibinfo{journal}{Pure and Applied Geophysics PAGEOPH}
  \bibinfo{volume}{125}, \bibinfo{pages}{777--812}.
\bibitem[{Saroglou and Tsiambaos(2008)}]{Saroglou2008}
\bibinfo{author}{Saroglou, H.}, \bibinfo{author}{Tsiambaos, G.},
  \bibinfo{year}{2008}.
\newblock \bibinfo{title}{{A modified Hoek-Brown failure criterion for
  anisotropic intact rock}}.
\newblock \bibinfo{journal}{International Journal of Rock Mechanics and Mining
  Sciences} \bibinfo{volume}{45}, \bibinfo{pages}{223--234}.
\bibitem[{Saussine et~al.(2006)Saussine, Cholet, Gautier, Dubois, Bohatier and
  Moreau}]{Saussine2006}
\bibinfo{author}{Saussine, G.}, \bibinfo{author}{Cholet, C.},
  \bibinfo{author}{Gautier, P.}, \bibinfo{author}{Dubois, F.},
  \bibinfo{author}{Bohatier, C.}, \bibinfo{author}{Moreau, J.},
  \bibinfo{year}{2006}.
\newblock \bibinfo{title}{{Modelling ballast behaviour under dynamic loading.
  Part 1: A 2D polygonal discrete element method approach}}.
\newblock \bibinfo{journal}{Computer Methods in Applied Mechanics and
  Engineering} \bibinfo{volume}{195}, \bibinfo{pages}{2841--2859}.
\bibitem[{Scholt{\`{e}}s and Donz{\'{e}}(2013)}]{Scholtes2013}
\bibinfo{author}{Scholt{\`{e}}s, L.}, \bibinfo{author}{Donz{\'{e}}, F.V.},
  \bibinfo{year}{2013}.
\newblock \bibinfo{title}{{A DEM model for soft and hard rocks: Role of grain
  interlocking on strength}}.
\newblock \bibinfo{journal}{J. Mech. Phys. Solids} \bibinfo{volume}{61},
  \bibinfo{pages}{352--369}.
\bibitem[{Sulem and Cerrolaza(2002)}]{Sulem2002}
\bibinfo{author}{Sulem, J.}, \bibinfo{author}{Cerrolaza, M.},
  \bibinfo{year}{2002}.
\newblock \bibinfo{title}{{Finite element analysis of the indentation test on
  rocks with microstructure}}.
\newblock \bibinfo{journal}{Computers and Geotechnics} \bibinfo{volume}{29},
  \bibinfo{pages}{95--117}.
\bibitem[{Turcotte(1986)}]{Turcotte1986}
\bibinfo{author}{Turcotte, L.}, \bibinfo{year}{1986}.
\newblock \bibinfo{title}{{Fractals and Fragmentation}}.
\newblock \bibinfo{journal}{Journal of Geophysical Research}
  \bibinfo{volume}{91}, \bibinfo{pages}{1921--1926}.
\bibitem[{Xu et~al.(2020)Xu, Gutierrez, He and Meng}]{Xu2020}
\bibinfo{author}{Xu, G.}, \bibinfo{author}{Gutierrez, M.}, \bibinfo{author}{He,
  C.}, \bibinfo{author}{Meng, W.}, \bibinfo{year}{2020}.
\newblock \bibinfo{title}{{Discrete element modeling of transversely isotropic
  rocks with non-continuous planar fabrics under Brazilian test}}.
\newblock \bibinfo{journal}{Acta Geotechnica} \bibinfo{volume}{7}.
\bibitem[{Zhang and Einstein(2000)}]{Zhang2000}
\bibinfo{author}{Zhang, L.}, \bibinfo{author}{Einstein, H.H.},
  \bibinfo{year}{2000}.
\newblock \bibinfo{title}{{Estimating the intensity of rock discontinuities}}.
\newblock \bibinfo{journal}{Int. J. Rock Mech. Min. Sci} \bibinfo{volume}{37},
  \bibinfo{pages}{819--837}.
\bibitem[{Zhang and Buscarnera(2018)}]{Zhang2018}
\bibinfo{author}{Zhang, Y.}, \bibinfo{author}{Buscarnera, G.},
  \bibinfo{year}{2018}.
\newblock \bibinfo{title}{{Breakage mechanics for granular materials in
  surface-reactive environments}}.
\newblock \bibinfo{journal}{J. Mech. Phys. Solids} \bibinfo{volume}{112},
  \bibinfo{pages}{89--108}.

\end{thebibliography}

\end{document}